\documentclass[12pt]{article}
\usepackage{epsfig}\parskip 5pt plus 1pt
\newcommand{\nn}{\nonumber}

\newcommand{\be}{\begin{equation}}
\newcommand{\ee}{\end{equation}}
\newcommand{\bea}{\begin{eqnarray}}
\newcommand{\eea}{\end{eqnarray}}
\begin{document}
\thispagestyle{empty}
\begin{flushright}
{\tt hep-ph/0206034}\\
{ROMA-1336/02}
\end{flushright}
\vspace*{1cm}
\begin{center}
{\Large{\bf The silver channel at the Neutrino Factory} }\\
\vspace{.5cm}
A. Donini$^{\rm a,}$\footnote{andrea.donini@roma1.infn.it},
D. Meloni$^{\rm b,}$\footnote{davide.meloni@roma1.infn.it}
and 
P. Migliozzi$^{\rm c,}$\footnote{pasquale.migliozzi@cern.ch}
 
\vspace*{1cm}
$^{\rm a}$ I.N.F.N., Sezione di Roma I and Dip. Fisica, 
Universit\`a di Roma ``La Sapienza'', 
P.le A. Moro 2, I-00185, Rome, Italy \\ 
$^{\rm b}$ Dip. Fisica, Universit\`a di Roma ``La Sapienza''
and I.N.F.N., Sezione di Roma I, P.le A. Moro 2, I-00185, Rome, Italy \\
$^{\rm c}$ I.N.F.N., Sezione di Napoli, Complesso Universitario di Monte Sant'Angelo, Via Cintia ed. G, I-80126 Naples, Italy 
\end{center}

\vspace{.3cm}
\begin{abstract}
\noindent
We notice that looking for $\nu_e \to \nu_\tau$
at the same time as $\nu_e \to \nu_\mu$ oscillations
could significantly help to reduce the errors in the 
leptonic CP-violating phase $\delta$ measurement.  
We show how the $\nu_e \to
\nu_\mu$ (``golden'') and $\nu_e \to \nu_\tau$ (``silver'')
transitions observed at an OPERA-like 2 Kton lead-emulsion detector at
$L = 732$ Km, in combination with the $\nu_e \to \nu_\mu$ transitions
observed at a 40 Kton magnetized iron detector with a baseline of $L =
3000$ Km, strongly reduce the so-called $(\theta_{13}, \delta)$
ambiguity. We also show how a moderate increase in the OPERA-like
detector mass (4 Kton instead of 2 Kton) completely eliminates
the clone regions even for small values of $\theta_{13}$.
\end{abstract}

\newpage

\section{Introduction}

The present atmospheric \cite{Fukuda:1994mc}-\cite{Toshito:2001dk} 
and solar \cite{Cleveland:1998nv}-\cite{Ahmad:2002ka} neutrino data 
are strongly supporting the hypothesis of neutrino oscillations 
\cite{Pontecorvo:1957yb}-\cite{Gribov:1969kq} 
and can be easily accommodated in a three family mixing scenario.

Let the Pontecorvo-Maki-Nakagawa-Sakata $U_{PMNS}$ matrix be 
the leptonic analogue of the hadronic Cabibbo-Kobayashi-Maskawa (CKM) 
mixing matrix in its most conventional parametrization \cite{PDG}:
\begin{eqnarray}
 U_{PMNS} =
\left ( 
\begin{array}{ccc}
1 &        0 &       0 \\
0 &   c_{23} &  s_{23} \\
0 & - s_{23} &  c_{23} 
\end{array}
\right ) \;
\left (
\begin{array}{ccc}
c_{13} & 0 &  s_{13} e^{i \delta} \\
0 & 1 & 0 \\
-  s_{13} e^{-i \delta} & 0 &  c_{13}
\end{array}
\right ) \;
\left (
\begin{array}{ccc}
 c_{12} &  s_{12} & 0 \\
- s_{12} &  c_{12} & 0 \\
0 & 0 & 1
\end{array}
\right ) \, , \nn \\
\nn
\end{eqnarray}
with the short-form notation $ s_{ij} \equiv \sin \theta_{ij}, 
c_{ij} \equiv \cos \theta_{ij}$.
Oscillation experiments are sensitive to the two neutrino mass differences $\Delta m^2_{12},
\Delta m^2_{23}$ and to the four parameters in the mixing matrix: three angles 
and the Dirac CP-violating phase, $\delta$. 

In particular, data on atmospheric neutrinos are interpreted as oscillations 
of muon neutrinos into neutrinos that are not $\nu_e$'s, with a mass gap that we denote 
by $\Delta m^2_{23}$. The corresponding mixing angle is close to maximal, 
$\sin^2 2 \theta_{23} > 0.8$, and $|\Delta m^2_{23}|$ is in the range 
$1.8$ to $4 \times 10^{-3}$ eV$^2$ \cite{Fogli:2001zx}.

The recent SNO results for solar neutrinos \cite{Ahmad:2001an}-\cite{Ahmad:2002ka}
favour the LMA-MSW \cite{MSW} solution of the solar neutrino deficit with 
$\nu_e$ oscillations into active ($\nu_\mu, \nu_\tau$) neutrino states.
The corresponding squared mass difference, that in this parametrization 
should be identified with $\Delta m^2_{12}$, is $\sim 10^{-5}$-$10^{-4}$ eV$^2$.
Comprehensive analyses of the solar neutrino data, however, 
do not exclude the LOW-MSW solution \cite{Barger:2002iv}-\cite{Holanda:2002pp}, 
with $\Delta m^2_{12} \sim 10^{-7}$ eV$^2$. In both cases, the corresponding
mixing angle ($\theta_{12}$) is large (albeit not maximal). 

Finally, the LSND data \cite{Athanassopoulos:1998pv,Aguilar:2001ty} 
would indicate a $\nu_\mu \to \nu_e$ oscillation with a third, 
very distinct, neutrino mass difference: $\Delta m_{LSND}^2 \sim 0.3 - 6\;{\rm eV}^2$. 
The LSND evidence in favour of neutrino oscillation has not been confirmed 
by other experiments so far \cite{Kleinfeller:2000em}; 
the MiniBooNE experiment \cite{Church:1997jc} will be able to do it in the near future
\cite{Sorel:2002hd}.
In the absence of an independent confirmation of the LSND evidence, 
we restrict ourselves to the three neutrino mixing scenario 
(the impact of a Neutrino Factory in the case of four neutrino mixing has
been discussed in full detail in \cite{Donini:1999jc}-\cite{Donini:2001xp}).

These oscillation signals will be confirmed in ongoing and planned 
atmospheric and solar neutrino experiments, as well as in long baseline ones, 
with  the latter being free of model-dependent estimations of neutrino fluxes.
There is a strong case for going further in the fundamental quest 
of the neutrino masses and mixing angles, as a necessary step to unravel the 
fundamental new scale(s) behind neutrino oscillations.
In particular, it is possible that in ten years from now no information whatsoever
will be at hand regarding the $\theta_{13}$ angle (the key between 
the atmospheric and solar neutrino realms, for which the present bound 
is $\sin^2 (2 \theta_{13}) \leq 1 \times 10^{-1}$, \cite{chooz})
and the leptonic CP violating phase $\delta$. 

An experimental set-up with the ambitious goal of precision
measurement of the whole three-neutrino mixing parameter space
is under study. This experimental programme consists of the development
of a ``Neutrino Factory'' (high-energy muons decaying in the straight section 
of a storage ring and producing a very pure and intense neutrino beam, 
\cite{Geer:1998iz,DeRujula:1998hd}) and of suitably optimized detectors located 
far away from the neutrino source. 
The effort to prepare such very long baseline neutrino experiments
will require a time period covering this and the beginning of the following 
decade. It is therefore of interest to look for the optimal conceivable 
factory-detectors combination. One of its main 
goals would be the discovery of leptonic CP violation and, possibly, 
its study \cite{Dick:1999ed}-\cite{Cervera:2000kp}.
Previous analyses \cite{Albright:2000xi}-\cite{Adams:2001tv} 
on the foreseeable outcome of experiments at a 
Neutrino Factory have shown that the determination of the 
two still unknown parameters in the three--neutrino mixing matrix, 
$\theta_{13}$ and $\delta$, will be possible (if the LMA-MSW solution of the
solar neutrino deficit is confirmed).
The most sensitive method to study these topics is to measure the
transition probabilities involving $\nu_e$ and $\bar \nu_e$, in particular 
$\nu_e(\bar \nu_e) \rightarrow \nu_\mu(\bar \nu_\mu)$. This is what is 
called  the ``golden measurement at the {\it neutrino factory}''. 
Such a facility is indeed unique in providing high energy and intense $\nu_e (\bar \nu_e)$ 
beams. Since these beams contain no $\bar \nu_\mu (\nu_\mu)$, the transitions
of interest can be measured by searching for ``wrong-sign'' muons: 
negative (positive) muons  appearing in a massive detector 
with good muon charge identification capabilities \cite{Cervera:2000kp}.

An incredible amount of work has been devoted to this topic in the last few years: 
we refer the interested reader to \cite{Koike:1999hf}-\cite{Gomez-Cadenas:2002fz} 
and to the refs. therein 
for an overview of the status-of-the-art in all its different aspects; 
we address to \cite{Barger:2000nf}-\cite{Huber:2002mx} and refs. therein 
for a comparison of the physics reach of a conventional (super)beam and 
of a Neutrino Factory;
eventually, we point out that in \cite{Zucchelli:2001gp} the idea of 
a $\bar \nu_e$ beam originating from $\beta$-decay (the so-called ``$\beta$-beam''), 
was advanced: it appears that the physics reach of such a beam is
complementary to that of a conventional superbeam \cite{Campanelli:2002sm}.

In \cite{Burguet-Castell:2001ez} it has been noticed that the probability 
$P_{\nu_\alpha \to \nu_\beta} (\bar \theta_{13},\bar \delta,E_\nu )$ for neutrinos 
at a fixed energy (and for a given baseline) computed for a given theoretical input pair 
$(\bar \theta_{13},\bar \delta)$ defines a continuous equiprobability curve in the 
($\theta_{13}, \delta$) plane. 
Therefore, for a fixed energy, a continuum of solutions reproduce the 
input probability. A second equiprobability curve is defined in this 
plane by the probability for antineutrinos at the same energy and with
the same input parameters, 
$P_{\bar \nu_\alpha \to \bar \nu_\beta} (\bar \theta_{13}, \bar \delta, E_\nu )$. 
The two equiprobability curves have, quite generally, two intersection points: 
the first of them at ($\bar \theta_{13},\bar \delta$), the second at a different
point ($\tilde \theta_{13},\tilde \delta$).
It is the intersection of equiprobability curves from the 
neutrinos and antineutrinos that resolves the continuum degeneracy of solutions in the 
($\theta_{13}, \delta$) plane, restricting the allowed values for $\theta_{13}$ and 
$\delta$ to the two regions around ($\bar \theta_{13},\bar \delta$) and 
($\tilde \theta_{13},\tilde \delta$). 
This second intersection, however, introduces an ambiguity in the measurement of the physical 
values of $\theta_{13}$ and $\delta$. Different proposals have been suggested
to solve this ambiguity: in \cite{Burguet-Castell:2001ez} the ambiguity is solved
by fitting at two different baselines at the same time; another possibility is an increase
in the energy resolution of the detector, \cite{Freund:2001ui,Rubbia:2001pk,Bueno:2001jd};
see also \cite{Kajita:2001sb}.

New degeneracies have later been noticed \cite{Barger:2001yr}, resulting
from our ignorance of the sign of the $\Delta m^2_{atm}$ squared mass difference
(by the time the Neutrino Factory will be operational)
and from the approximate $[\theta_{23}, \pi/2 - \theta_{23}]$ symmetry for the atmospheric
angle.

In the first part of the paper we describe how the ($\theta_{13}, \delta$) ambiguity
arises in $\nu_e \to \nu_\mu$ oscillation due to the equiprobability curves 
in the ($\theta_{13},\delta$) plane at fixed neutrino energy. 
We then extend our analysis showing how the same phenomenon
can be observed in a real experiment: equal-number-of-events (ENE) curves 
for any given neutrino energy bin appears and their intersections in the 
($\theta_{13},\delta$) plane explain how and where do ``clone'' regions
arise. Our analysis is afterwards compared with the results of simulations 
with a realistic magnetized iron detector (studied in \cite{Cervera:2000vy}).

We then propose to reduce the continuum degeneracy and to resolve the
last ambiguity using two baselines with detectors of different design.
We notice that muons proceeding from $\tau$ decay when $\tau$'s are
produced via a $\nu_e \to \nu_\tau$ transition show a different
$(\theta_{13}, \delta)$ correlation from those coming from $\nu_e \to
\nu_\mu$ (first considered in \cite{Cervera:2000kp}).  By using a near
lead-emulsion detector, capable of the $\tau$-decay vertex
recognition, we can therefore use the complementarity of the
information from $\nu_e \to \nu_\tau$ and from $\nu_e \to \nu_\mu$ to
solve the ($\theta_{13}, \delta$) ambiguity.  We find that the
combination of the near emulsion detector and of a massive magnetized
iron detector at $L = 3000$ Km could indeed help to achieve a good
resolution in the $(\theta_{13}, \delta)$ plane.

In this paper we restrict ourselves to the ($\theta_{13},\delta$)
ambiguity, by fixing $\theta_{23} = 45^\circ$ and by choosing a given
sign for $\Delta m^2_{atm}$ (in the hypothesis that more information
on the three neutrino spectrum will be available by the time the
Neutrino Factory will be operational). However, we found that
the core of our results do not depend on the sign of $\Delta m^2_{atm}$: 
indeed, we have been carrying on simulations with the opposite sign, 
with similar results (i.e., we still observe how ``clone'' regions
disappear due to the two-detector types combination; notice, however,
that the location of the ``clone'' regions and all the 
details of the simulation do depend on the sign of $\Delta m^2_{atm}$).
We remind that, if $\theta_{13}$ is not extremely small, 
the combined measurement of $\nu_e \to \nu_\mu$ and $\nu_e \to \nu_\tau$
transitions could significantly help in solving the $[\theta_{23},\pi/2 - \theta_{23}]$
ambiguity. 

In Section \ref{sec:emuprob} we present our analysis of the $\nu_e \to
\nu_\mu$ equiprobability curves in the ($\theta_{13},\delta$) plane;
in Section \ref{sec:emuENE} we introduce the corresponding
equal-number-of-events (ENE) curves; in Section \ref{sec:etau} we
present a similar analysis for the $\nu_e \to \nu_\tau$ oscillation
probability and study the impact of ``silver'' wrong-sign muon events;
in Section \ref{sec:etauRES} we show our results for the combination
of a near ($L = 732$ Km) OPERA-like detector and of a $L = 3000$ Km
magnetized iron detector; in Section \ref{sec:concl} we eventually
draw our conclusions.  In Appendix \ref{app:dtheta} a perturbative
expansion in $\Delta \theta$ of the formulae of
Sect.~\ref{sec:emuprob} is presented; in Appendix \ref{app:form} we
report some useful formulae for $\tau$ CC-interaction and
decay.

\section{$\nu_e \to \nu_\mu$ equiprobability curves in the $(\theta_{13},\delta)$ plane}
\label{sec:emuprob}

We consider the $\nu_e \to \nu_\mu$ transition, first. This channel
has been shown to be the optimal one to measure simultaneously
$\theta_{13}$ and $\delta$ at the Neutrino Factory in the context of
three-family mixing, through the appearance of ``wrong-sign'' muons in
the detector, \cite{Cervera:2000kp}. It therefore deserves the
nickname of ``golden channel''.

Following eq.~(1) of \cite{Burguet-Castell:2001ez} we get for the
transition probability at second order in perturbation theory in
$\theta_{13}$, $\Delta_\odot/\Delta_{atm}$, $\Delta_\odot/A$ and
$\Delta_\odot L$ (see also
\cite{Freund:2001pn}-\cite{Minakata:2002qe}), \be
\label{eq:spagnoli}
P^\pm_{e \mu} (\bar \theta_{13}, \bar \delta) = 
X_\pm \sin^2 (2 \bar \theta_{13}) + 
Y_\pm \cos ( \bar \theta_{13} ) \sin (2 \bar \theta_{13} )
      \cos \left ( \pm \bar \delta - \frac{\Delta_{atm} L }{2} \right ) + Z \, ,
\ee
where $\pm$ refers to neutrinos and antineutrinos, respectively,
and
\be
\label{eq:emucoeff}
\left \{ 
\begin{array}{lll}
X_\pm &=& \sin^2 (\theta_{23} ) 
\left ( \frac{\Delta_{atm} }{ B_\mp } \right )^2
\sin^2 \left ( \frac{ B_\mp L}{ 2 } \right ) \ , \\
\nn \\
Y_\pm &=& \sin ( 2 \theta_{12} ) \sin ( 2 \theta_{23} )
\left ( \frac{\Delta_\odot }{ A } \right )
\left ( \frac{\Delta_{atm} }{ B_\mp } \right )
\sin \left ( \frac{A L }{ 2 } \right )
\sin \left ( \frac{ B_\mp L }{ 2 } \right ) \ , \\
\nn \\
Z &=& \cos^2 (\theta_{23} ) \sin^2 (2 \theta_{12})
\left ( \frac{\Delta_\odot }{ A } \right )^2
\sin^2 \left ( \frac{A L }{ 2 } \right ) \ ,
\end{array}
\right .  \ee with $Z = Z_+ = Z_-$. In these formulae, $A = \sqrt{2}
G_F n_e$ (expressed in eV$^2$/GeV) and $B_\mp = | A \mp \Delta_{atm}
|$ (with $\mp$ referring to neutrinos and antineutrinos,
respectively). Finally, $\Delta_{atm} = \Delta m^2_{atm} / 2 E_\nu$
and $\Delta_\odot = \Delta m^2_\odot / 2 E_\nu$.

The parameters $\bar \theta_{13}$ and $\bar \delta$ are the physical
parameters that must be reconstructed by fitting the experimental data
with the theoretical formula for oscillations in matter.  In what
follows, the other parameters have been considered as fixed
quantities, supposed to be known by the time when the Neutrino Factory
will be operational.  In particular, in the solar sector we fixed
$\theta_{12} = 33^\circ$ and $\Delta m^2_\odot = 1.0 \times 10^{-4} $
eV$^2$ \cite{Barger:2002iv}-\cite{Holanda:2002pp}, corresponding to
the LMA region of the solar neutrino problem (accordingly to the
recent SNO results \cite{Ahmad:2002jz,Ahmad:2002ka}); in the
atmospheric sector, $\theta_{23} = 45^\circ$ and $\Delta m^2_{atm} =
2.9 \times 10^{-3} $ eV$^2$ \cite{Fogli:2001zx}
(notice that for $\theta_{23} = 45^\circ$ the $[\theta_{23},\pi/2 -
\theta_{23}]$ ambiguity \cite{Barger:2001yr} is absent).
Finally, we considered a fixed value for the matter parameter, $A =
1.1 \times 10^{-4} $ eV$^2$/GeV for $L < 4000$ km and $A = 1.5 \times
10^{-4} $ eV$^2$/GeV for $L > 4000$ km, obtained by using the average
matter density alongside the path for the chosen distance computed
with the Preliminary Earth Model \cite{earthmodel}.
For simplicity, we have not included errors on these
parameters\footnote{ It has been shown in
\cite{Burguet-Castell:2001ez} that the inclusion of the foreseeable
uncertainties on these parameters does not modify the results on the
$\theta_{13}$ and $\delta$ measurements in a relevant manner.}.

Eq.~(\ref{eq:spagnoli}) leads to an equiprobability curve in the plane 
($\theta_{13}, \delta$) for neutrinos and antineutrinos of a given energy:
\be
P^\pm_{e \mu} (\bar \theta_{13}, \bar \delta) = P^\pm_{e \mu} (\theta_{13}, \delta) \, .
\label{eq:equi0}
\ee
We can solve eq.~(\ref{eq:equi0}) for $\delta$: 
\be
\cos \left ( \pm \delta - \frac{\Delta_{atm} L}{2} \right ) =
\frac{ \bar P^\pm_{e\mu} - X_\pm \sin^2 (2 \theta_{13} ) - Z}{
Y_\pm \cos ( \theta_{13}) \sin (2 \theta_{13} ) } \, .
\label{eq:equi1}
\ee
It is useful to introduce the following functions: 
\be
\left \{ \begin{array}{ccc}
          f (\theta_{13}, \bar \theta_{13}) &=& 
                  \frac{\sin^2 (2 \bar \theta_{13}) - \sin^2 (2 \theta_{13}) }{ 
                                       \cos \theta_{13} \sin (2 \theta_{13}) } \, , \\
\\
          g (\theta_{13}, \bar \theta_{13}) &=& 
                       \frac{\cos \bar \theta_{13} \sin (2 \bar \theta_{13}) }{ 
                                       \cos \theta_{13} \sin (2 \theta_{13}) } \, ,
         \end{array} \right .
\label{eq:functions}
\ee
with the obvious limit $f(\bar \theta_{13}, \bar \theta_{13}) = 0$ 
and $g (\bar \theta_{13}, \bar \theta_{13}) = 1$.

\noindent Eq.~(\ref{eq:equi1}) can then be written as:  
\be
\cos \left (\pm \delta - \frac{\Delta_{atm} L}{2} \right ) = 
R_\pm f (\theta_{13}, \bar \theta_{13}) + 
\cos \left ( \pm \bar \delta - \frac{\Delta_{atm} L}{2} \right ) 
g (\theta_{13}, \bar \theta_{13}) \, . 
\label{eq:equi2}
\ee Eq.~(\ref{eq:equi2}) is particularly illuminating: it describes a
family of two branches curves in the plane ($\theta_{13},\delta$) for
the neutrinos and a second family of two branches curves for the
antineutrinos. The dependence on the neutrino energy resides in
$\Delta_{atm}$ and in the ratio $R_\pm = X_\pm / Y_\pm$, whereas the
dependence on the angle is factorized in the two
$\theta_{13}$-dependent functions $f$ and $g$.

\begin{figure}[h!]
\begin{center}
\begin{tabular}{ccc}
\hspace{-1cm} \epsfxsize5cm\epsffile{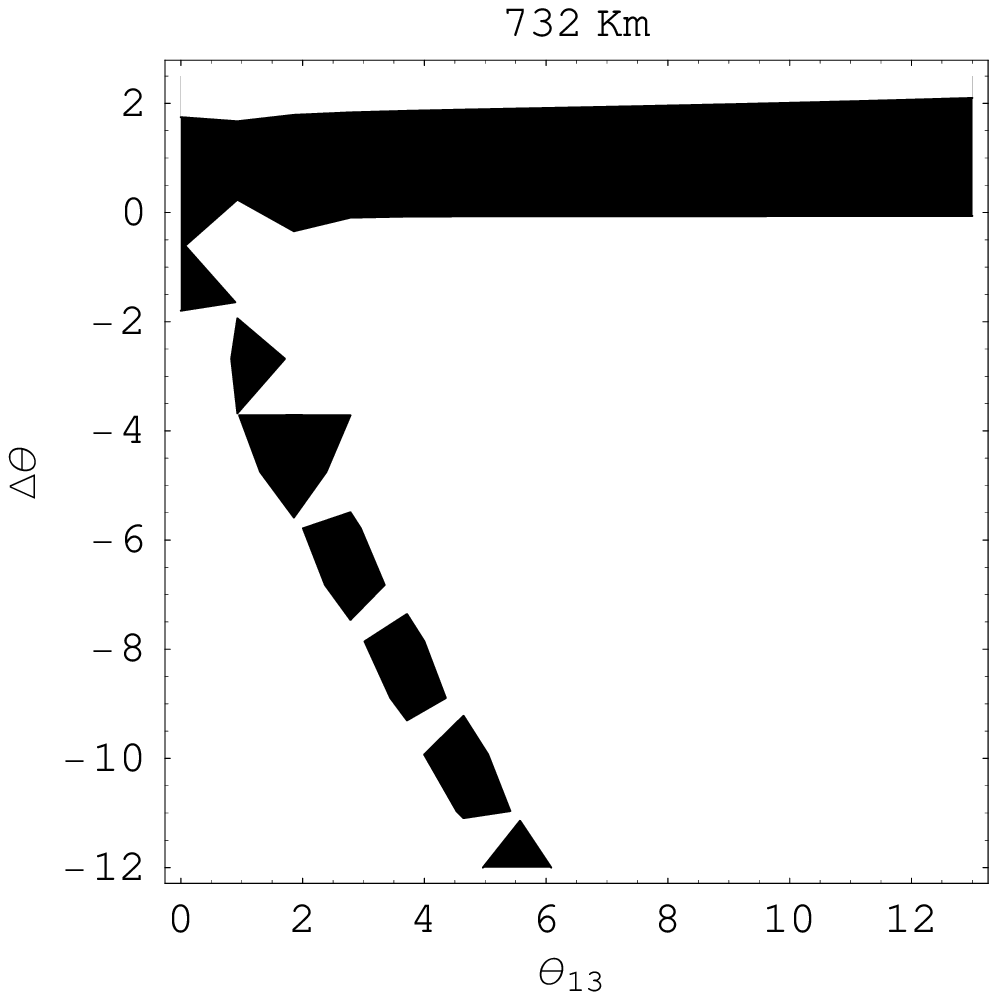} &
              \epsfxsize5cm\epsffile{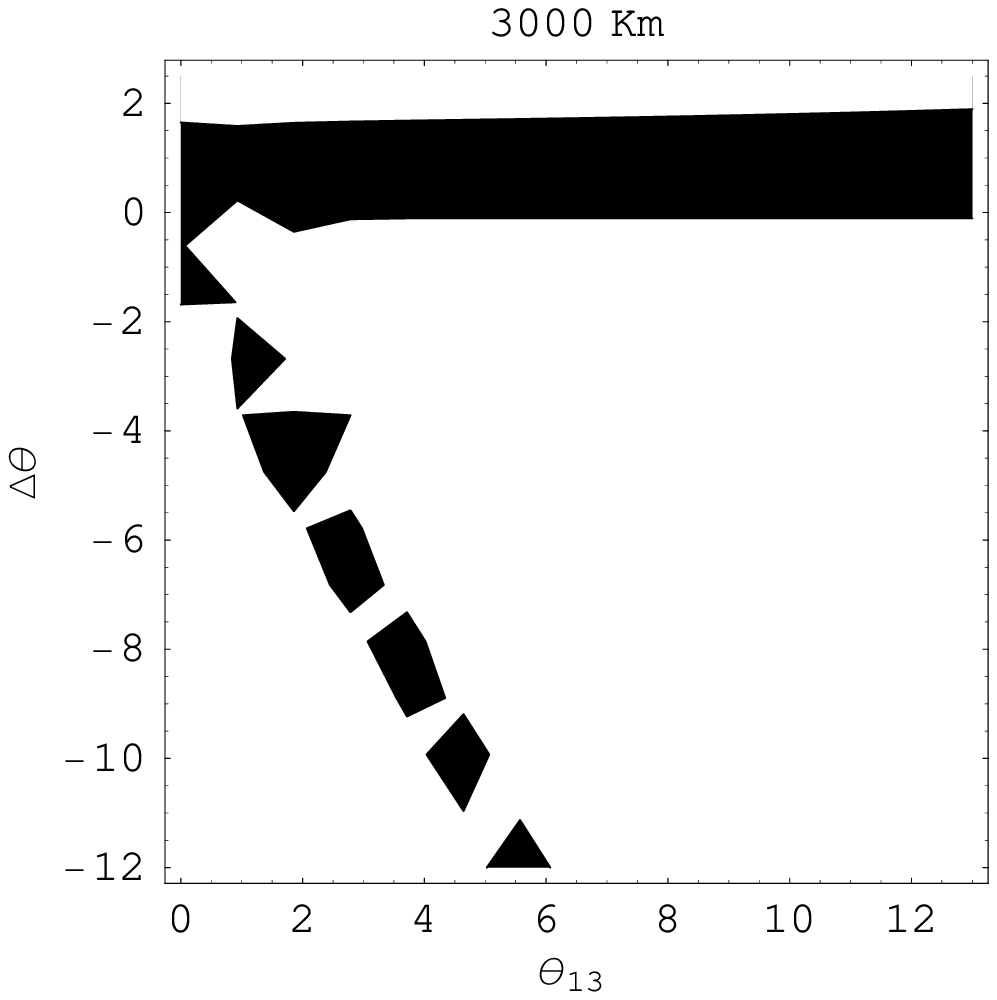} &
              \epsfxsize5cm\epsffile{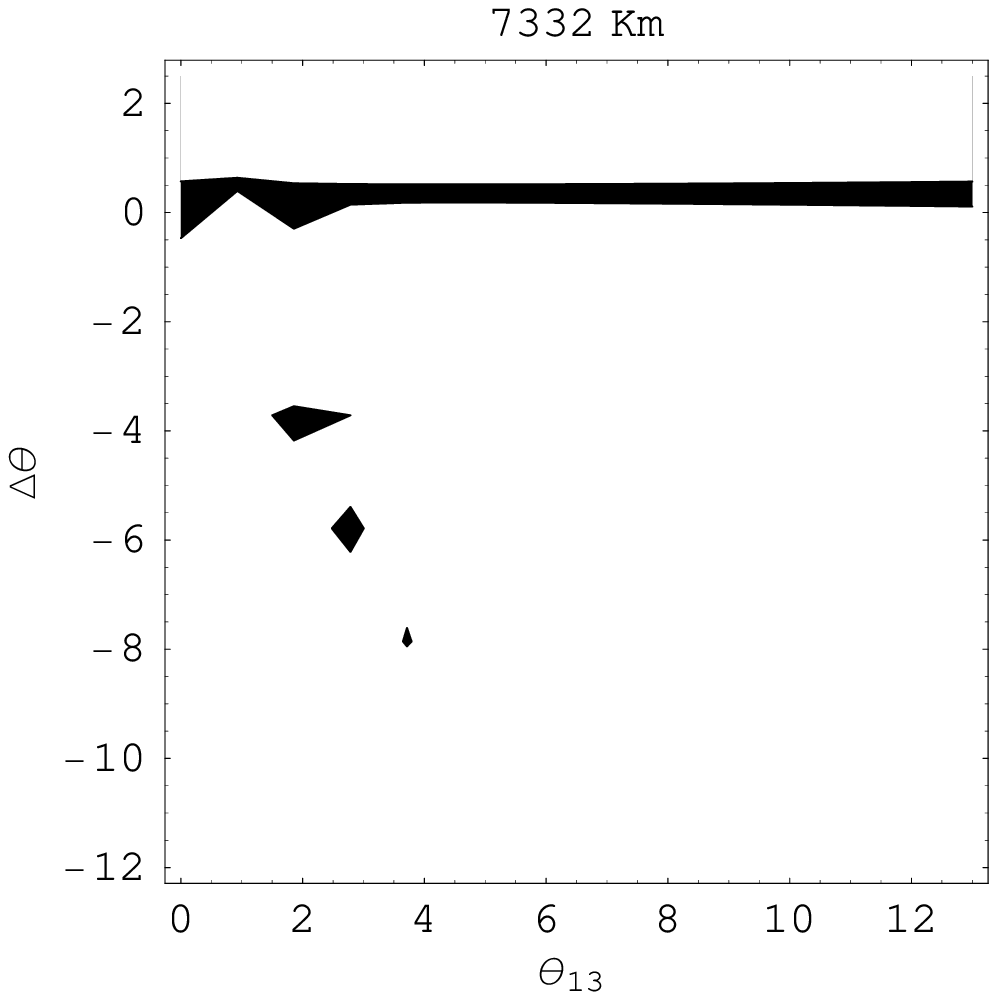}
\end{tabular}
\caption{\it Allowed region in $\Delta \theta$ as a function of $\theta_{13}$
for $E_\nu =38$ GeV at three different distances: $L = 732, 3000$ and $7332$ Km.}
\label{fig:dthetalim}
\end{center}
\end{figure}

\noindent It is helpful to introduce the parameter $\Delta \theta$:
\bea
\theta_{13} = \bar \theta_{13} + \Delta \theta \, , \nn \\
\nn
\eea constrained by the bound
\be
\left | \cos\left ( \pm \delta - \frac{\Delta_{atm} L}{2} \right ) \right | \leq 1 \, .
\label{eq:bound}
\ee The allowed region for $\Delta \theta$ depend on the input
parameters ($\bar \theta_{13}, \bar \delta$), on the neutrino energy
and on the baseline.  In Fig.~\ref{fig:dthetalim}, we compute the
allowed values of $\Delta \theta$ as a function of $\bar \theta_{13}$
($ \bar \theta_{13} \in [0^\circ, 13^\circ]$) at three different
baselines, $L = 732, 3000$ and $7332$ Km, for $\bar \delta = 0$ and
$E_\nu = 38$ GeV, by numerically solving eq.~(\ref{eq:bound}).

It may be noticed that, for (almost) every value of $\bar \theta_{13}$
in the considered range, two different regions of allowed values for
$\Delta \theta$ exist. The first region corresponds to $\Delta \theta
\simeq 0$, whereas the second corresponds to large negative values
for $\Delta \theta$. In this region, $\theta_{13}$ is negative: this
region is therefore unphysical when the sign of the mass differences
and of the various angles are defined in an appropriate way
\cite{Lipari:2001ds}.  We concentrate hereafter on the tiny region
around $\Delta \theta \simeq 0$.

\begin{figure}[h!]
\begin{center}
\begin{tabular}{ccc}
\hspace{-1cm} \epsfxsize5cm\epsffile{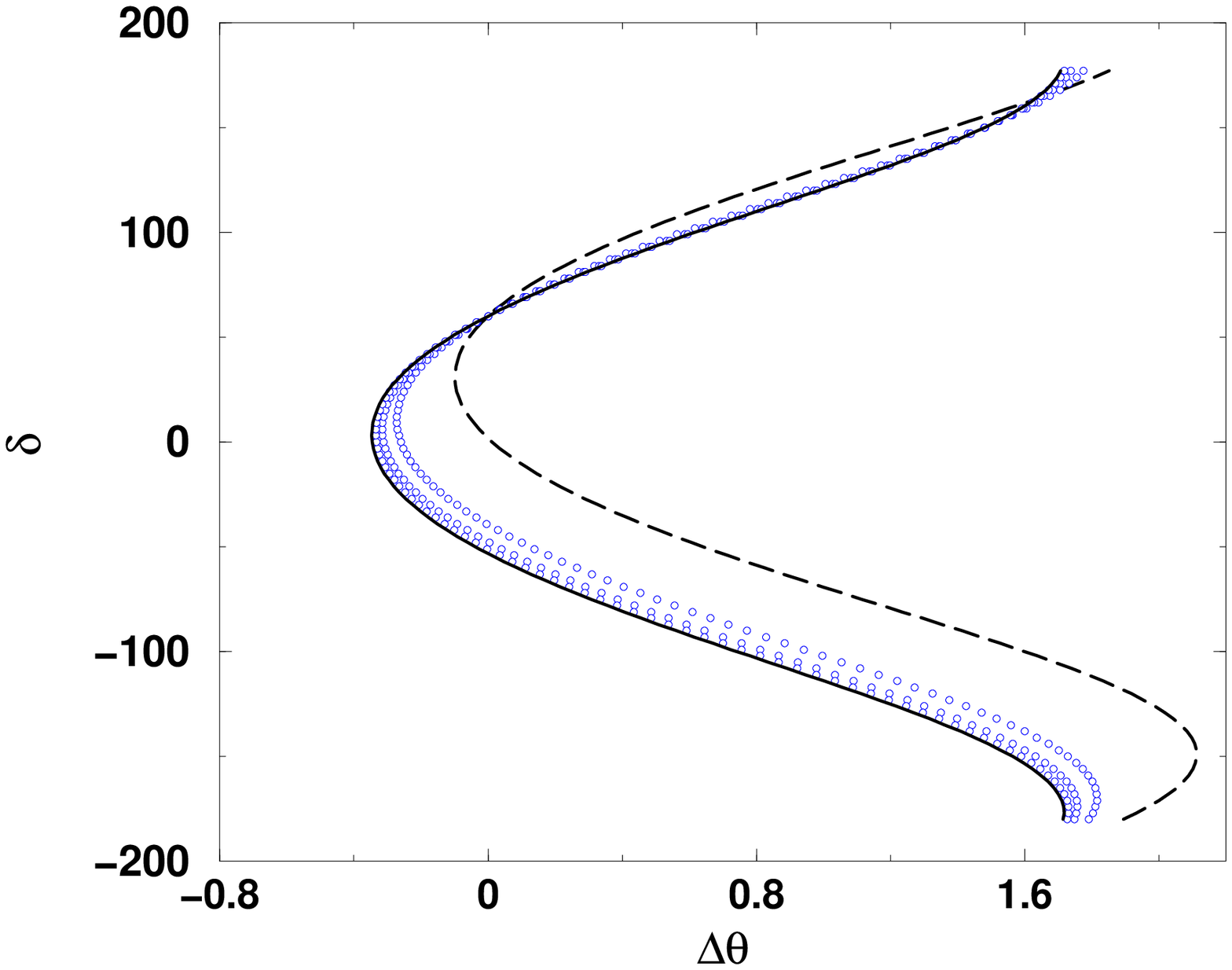} &
              \epsfxsize5cm\epsffile{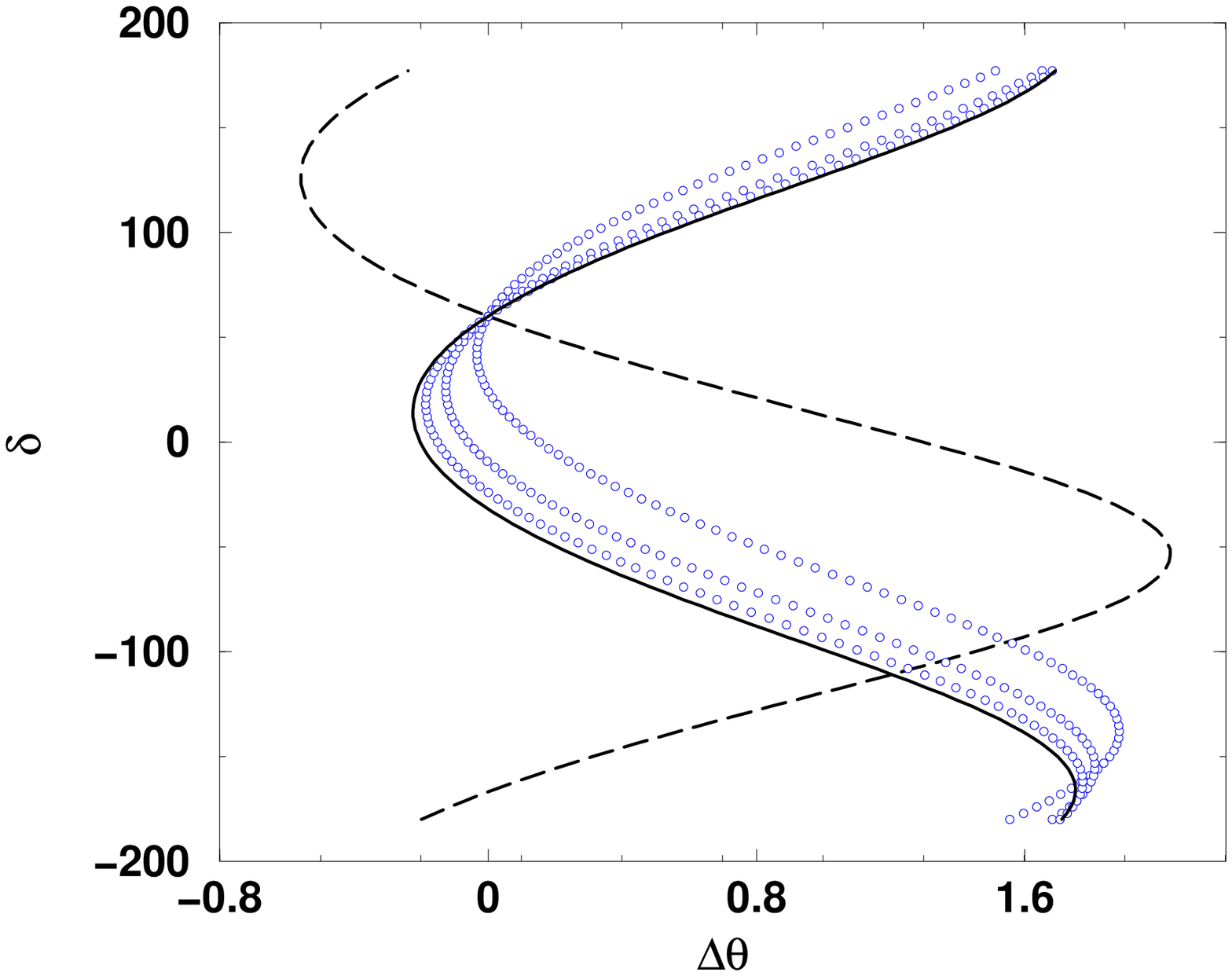} &
              \epsfxsize5cm\epsffile{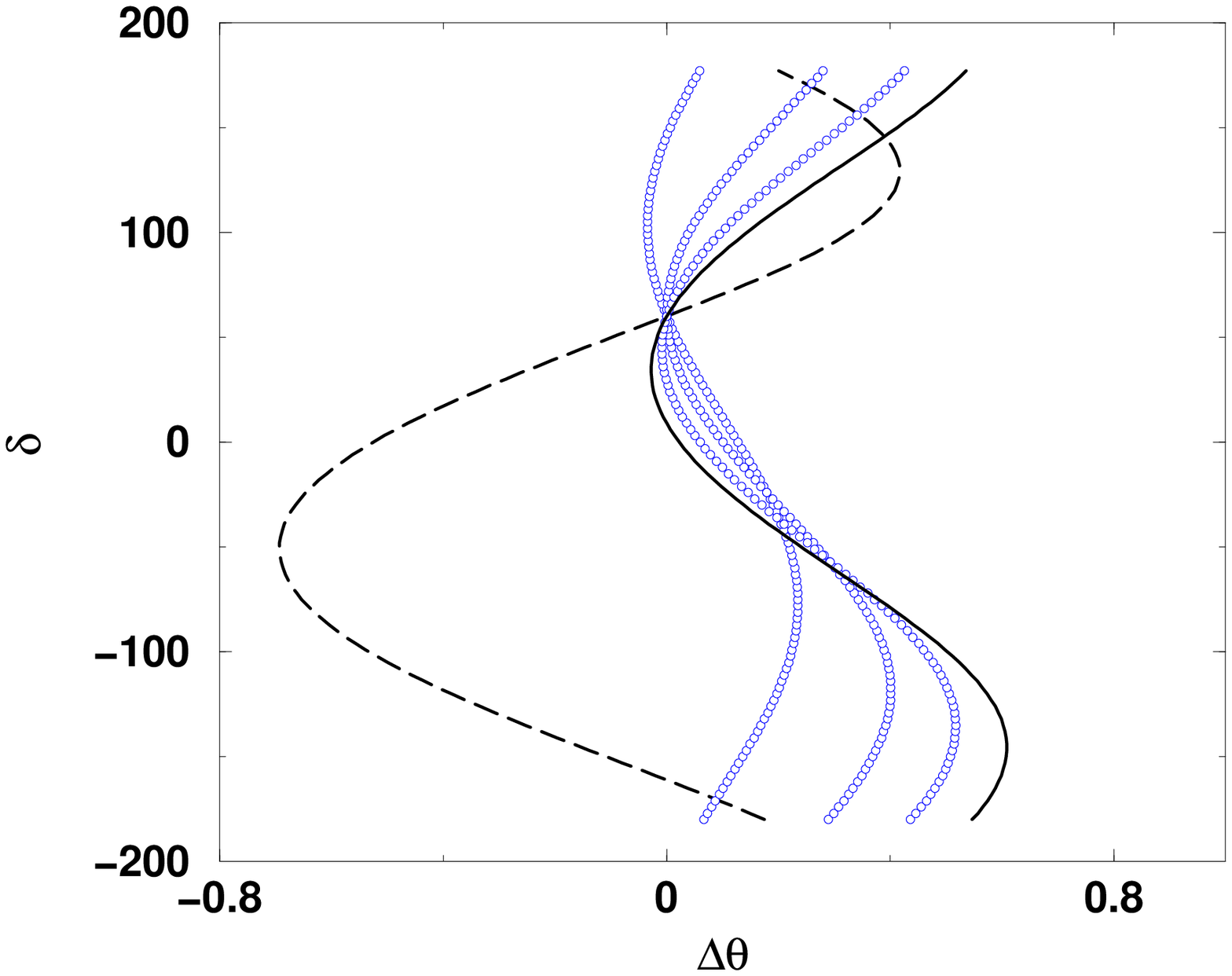} \\
\hspace{-1cm} \epsfxsize5cm\epsffile{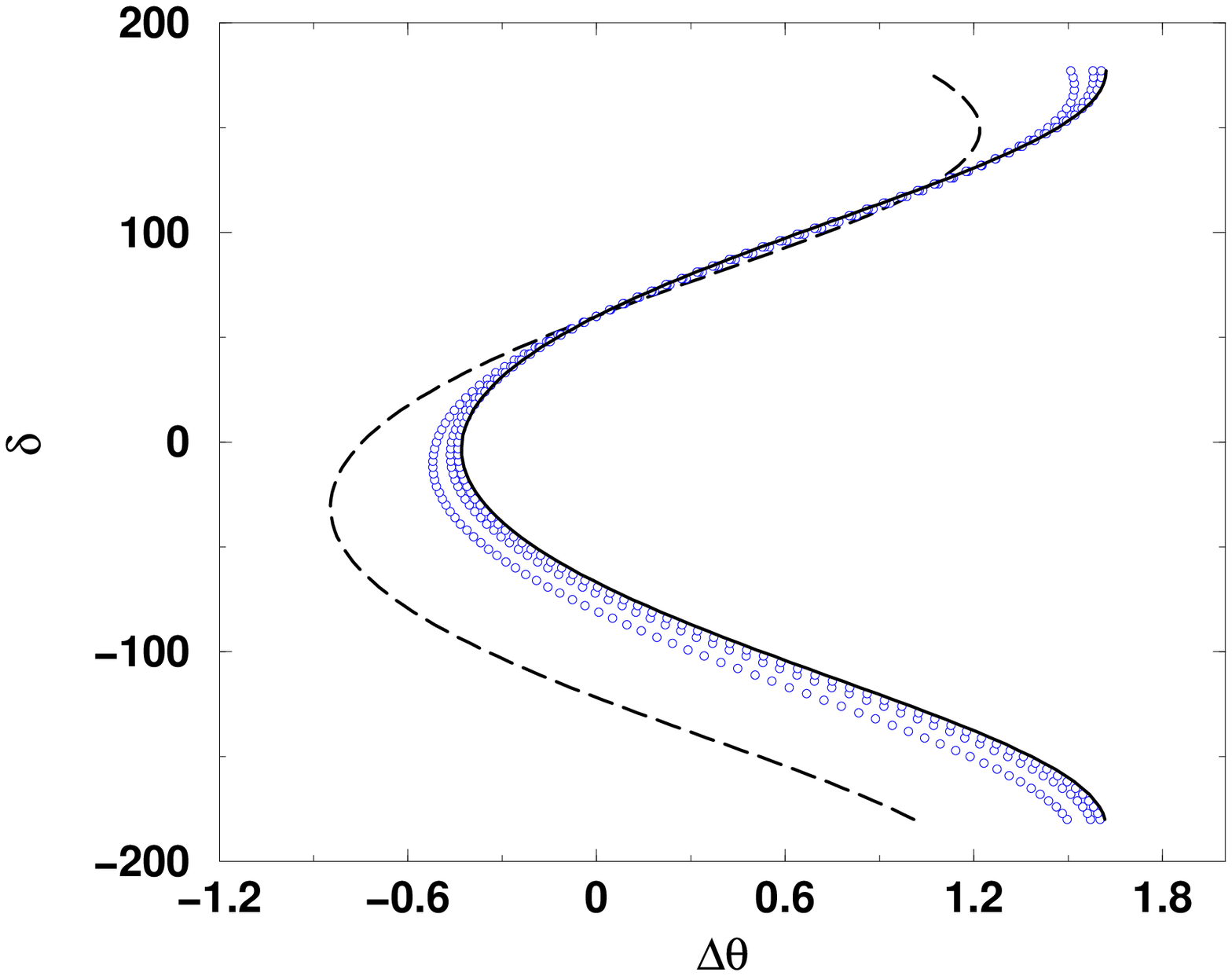} &
              \epsfxsize5cm\epsffile{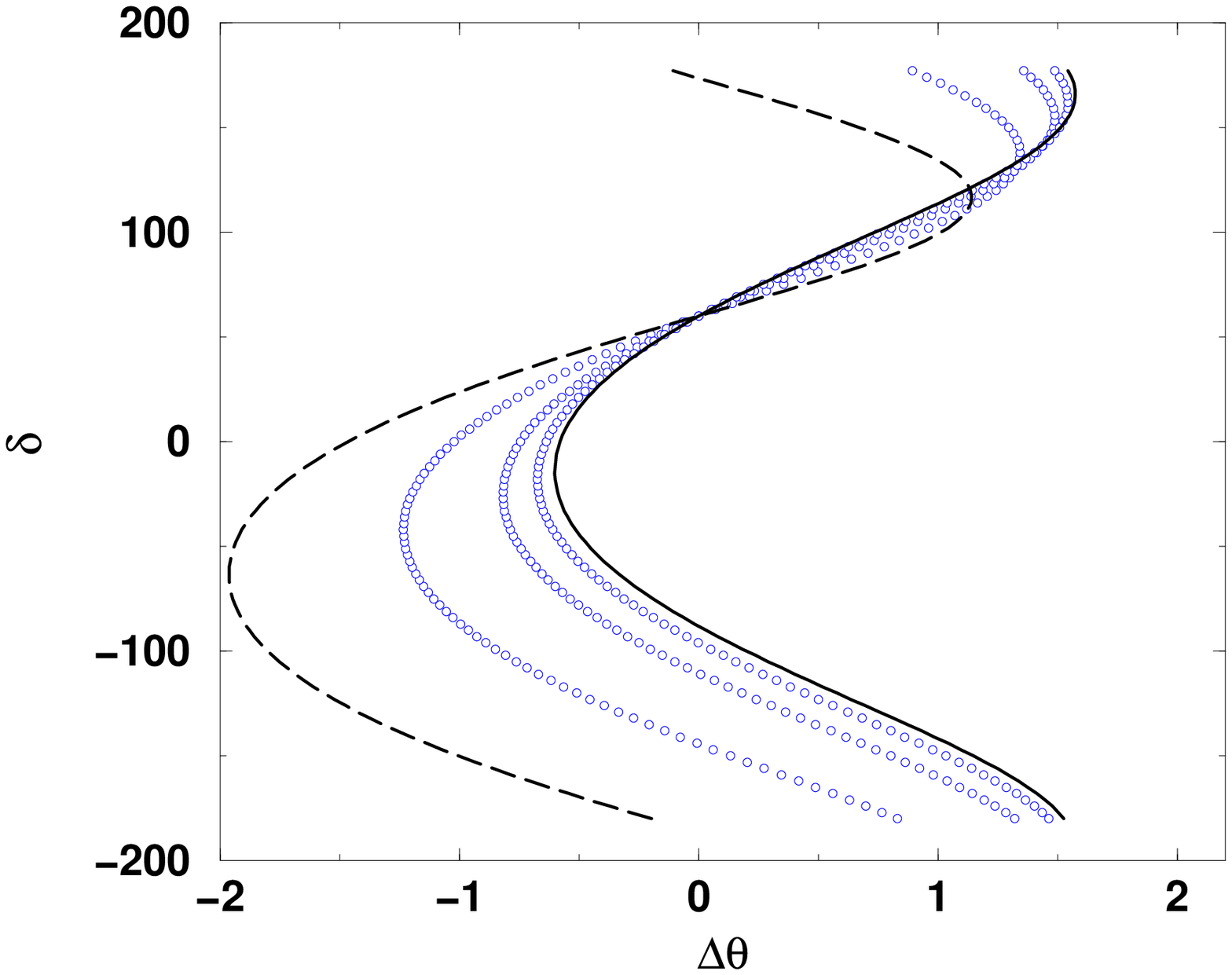} &
              \epsfxsize5cm\epsffile{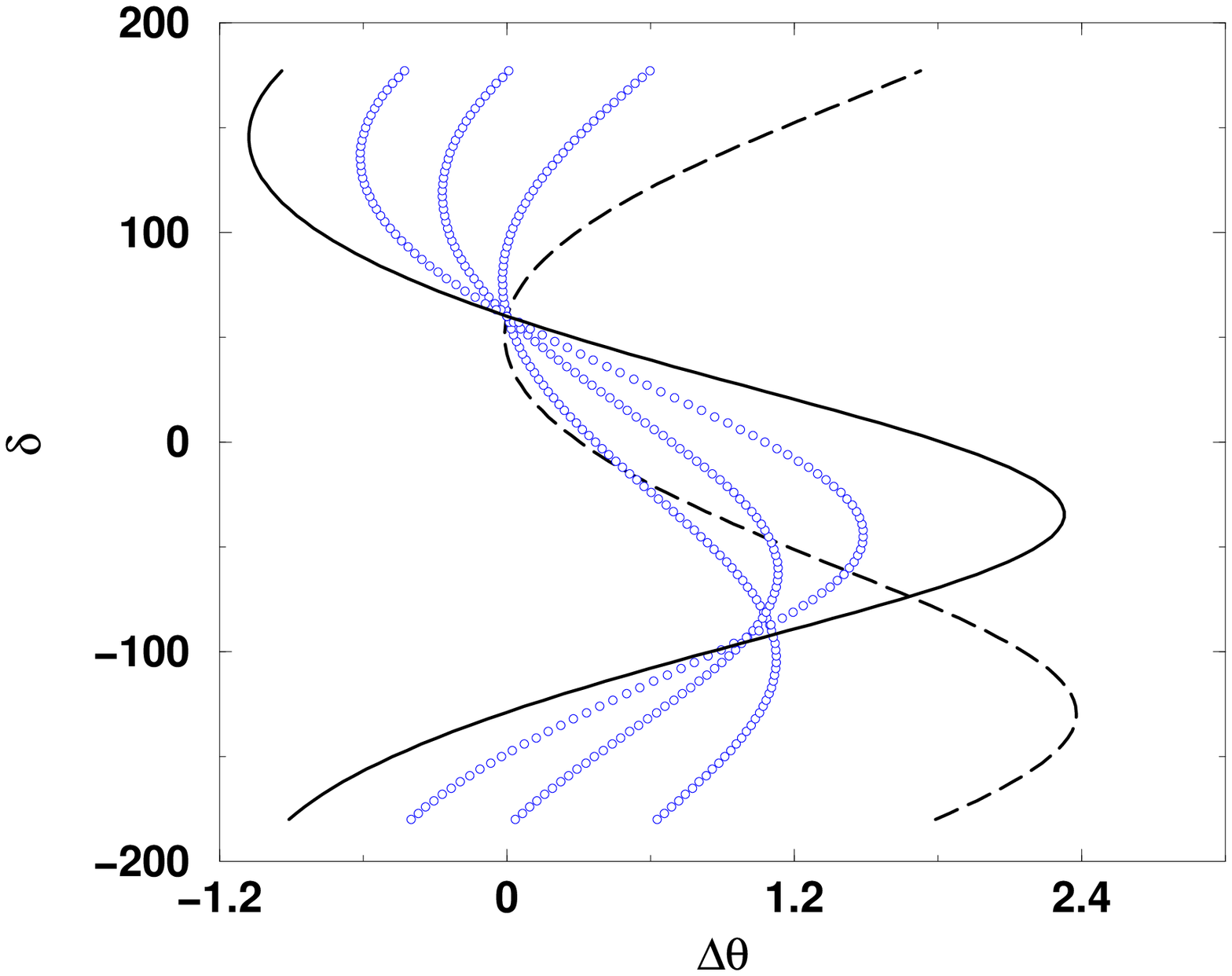}
\end{tabular}
\caption{\it Equiprobability curves in the ($\Delta \theta, \delta$)
plane, for $\bar \theta_{13} = 5^\circ, \bar \delta = 60^\circ$,
$E_\nu \in [5, 50] $ GeV and $L = 732, 3000$ and $7332$ Km. The upper
row represents equiprobability curves for neutrinos, the lower row for
antineutrinos. The dashed line is $E_\nu = 5$ GeV, the solid line is
$E_\nu = 45$ GeV; the dotted lines lie in between these two.}
\label{fig:equiprob}
\end{center}
\end{figure}

In Fig.~\ref{fig:equiprob} we present the equiprobability curves for
$P^+_{e\mu}$ (the upper row) and $P^-_{e\mu}$ (the lower row) in the
($\Delta \theta, \delta$) plane, at $L = 732, 3000$ and $7332$ Km for
different values of the neutrino energy in the range $E_\nu \in [5,
50]$ GeV. The input values are $\bar \theta_{13} = 5^\circ$ and $\bar
\delta = 60^\circ$.  In the upper row (neutrinos), it can be seen that
all the equiprobability curves intersect in $\Delta \theta = 0^\circ,
\delta = 60^\circ$ (namely, $\theta_{13} = \bar \theta_{13}$ and
$\delta = \bar \delta$).  However, notice that the equiprobability
curve for a given neutrino energy intersects the curves corresponding
to a different neutrino energy in a second point, at positive $\Delta
\theta$ and negative $\delta$. This second intersection depends on the
energies of the two curves.  In the upper branch of the neutrino
equiprobability curves, no second intersection is observed, for these
particular values of the input parameters. 
The results of Fig.~\ref{fig:equiprob} may be understood with the
help of a perturbative expansion of eqs.~(\ref{eq:functions})
and (\ref{eq:equi2}) in terms of powers of $\Delta \theta$ 
(always possible in the allowed region, for $\bar \theta_{13}$ large enough).
Details on this expansion can be found in App.~\ref{app:dtheta}.

We can draw some conclusion from what observed in
Fig.~\ref{fig:equiprob} and from the previous considerations on the
energy dependence of the equiprobability curves.  In particular, it is
to be expected that by fitting experimental data for neutrinos only it
should be quite difficult to determine the physical parameters ($\bar
\theta_{13}, \bar \delta$) with good accuracy. We expect, instead,
that the fitting procedure will identify a low $\chi^2$ region
whenever the family of equiprobability curves are not well separated,
within the experimental energy resolution.  In particular, at short
distance ($L = 732$ Km) it is to be expected a good determination of
$\bar \theta_{13}$ (notice that $\Delta \theta$ is generally less than
$2^\circ$) and no determination whatsoever of the CP-violating phase
$\bar \delta$. At the intermediate distance, $L = 3000$ Km, the
equiprobability curves for neutrinos (for this particular set of
input parameters) do not depend strongly on the energy in the upper branch, 
whereas a larger separation can be seen in the lower branch. 
Therefore, we expect a low $\chi^2$ region alongside
the upper branch of the equiprobability curves spanning from around
the single point corresponding to the physical parameters (at $\Delta
\theta = 0, \delta = \bar \delta$) to the (diluted) region where the
curves show the second intersection (that by periodicity in the $\delta$
axis happens to be in the lower branch).

Finally, at large distance ($L = 7332$ Km) 
we expect a low $\chi^2$ region in the region $\Delta \theta
\leq 0.5^\circ$ and $\delta \simeq \bar \delta \pm 100^\circ$ (notice
that the small spread in the variable $\theta$ for this baseline is in
agreement with Fig.~\ref{fig:dthetalim}).

\begin{figure}[h!]
\begin{center}
\begin{tabular}{ccc}
\hspace{-1cm} \epsfxsize5cm\epsffile{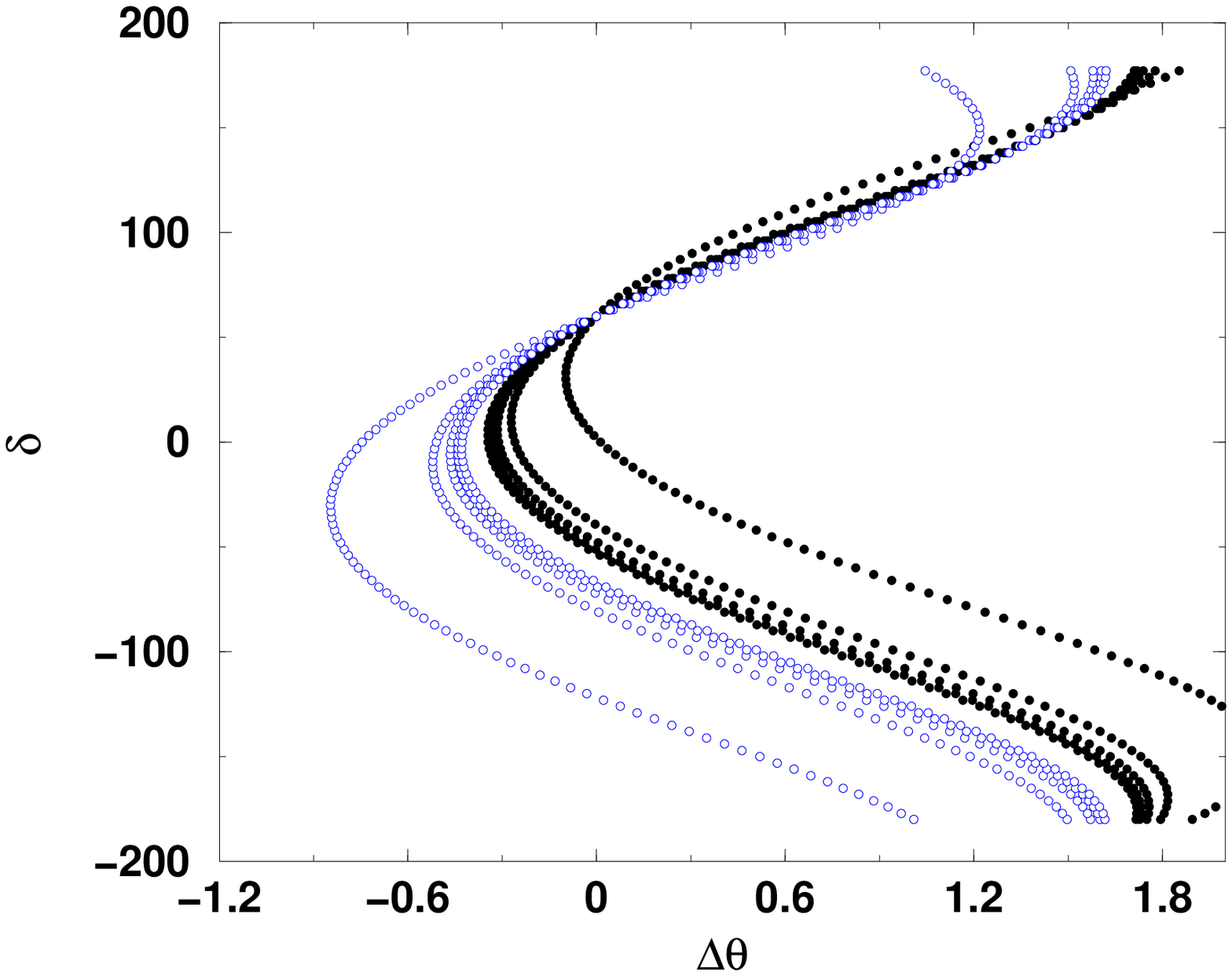} &
              \epsfxsize5cm\epsffile{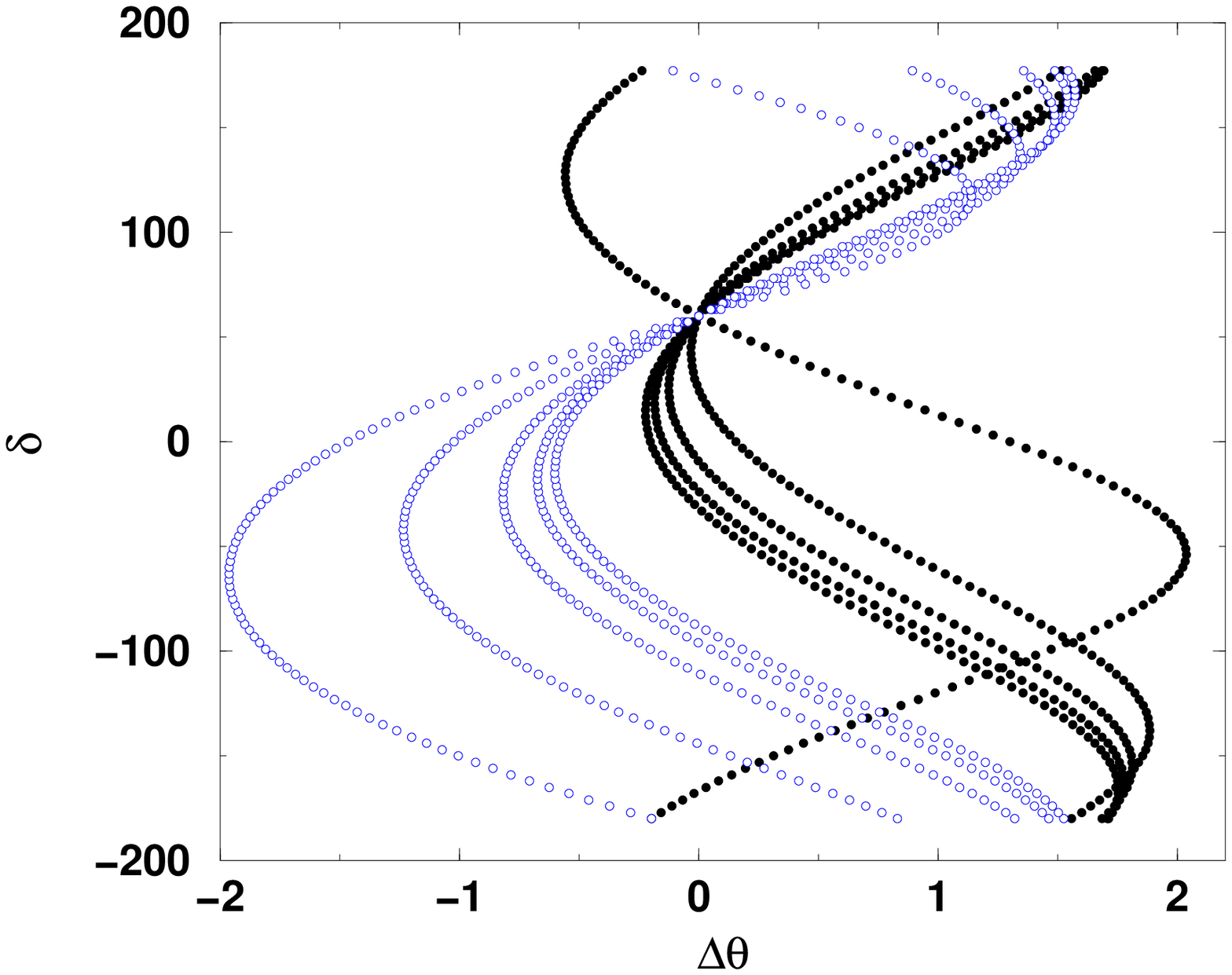} &
              \epsfxsize5cm\epsffile{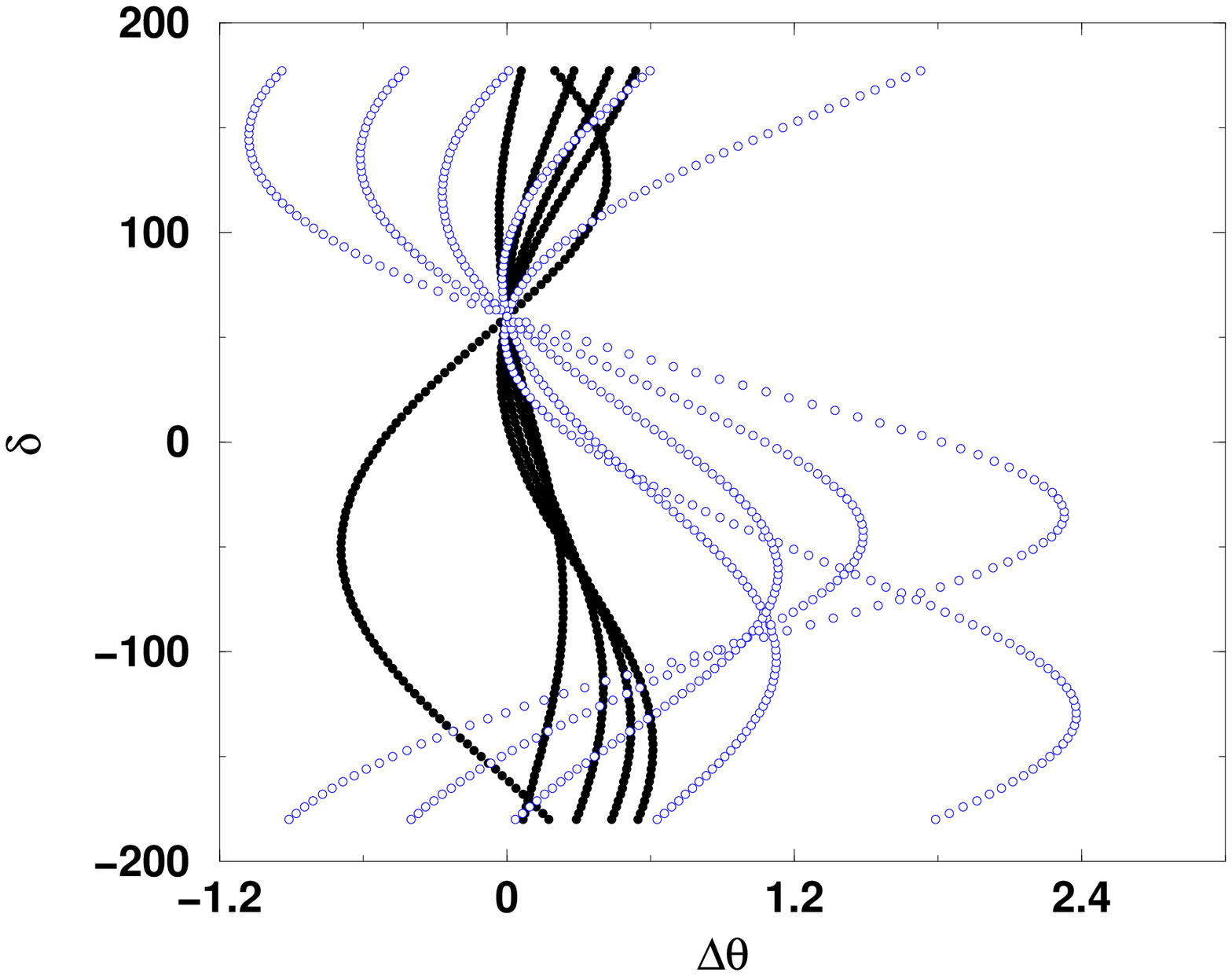}
\end{tabular}
\caption{\it Same as in Fig.~\ref{fig:equiprob}, but with neutrinos and antineutrinos
equiprobability curves superimposed.}
\label{fig:allprob}
\end{center}
\end{figure}

A great improvement in the reconstruction of the physical parameters
from the experimental data is achievable using at the same time
neutrino and antineutrino data. This can be seen in
Fig.~\ref{fig:allprob}, where the equiprobability curves for neutrinos
and antineutrinos (for the same input parameters as in
Fig.~\ref{fig:equiprob}) have been superimposed. At short distance,
the two family of equiprobability curves overlap for any value of
$\delta$, and no improvement is to be expected. However, at the
intermediate distance the equiprobability curves for neutrinos and
antineutrinos overlap only in the vicinity of the physical point
$\Delta \theta = 0, \delta = \bar \delta$ and in the region of the
second intersection, whereas in the intermediate region they are quite
well separated, both in the upper and lower branch. We expect, in this
case, that the fitting procedure of the whole set of neutrino and
antineutrino data will identify two separate low $\chi^2$ region,
around the physical point and around the region where all the curves
show the second intersection.  This second allowed region in the
parameter space was first observed in \cite{Burguet-Castell:2001ez}
and subsequently confirmed in
\cite{Freund:2001ui,Rubbia:2001pk,Bueno:2001jd}.  Finally, at large
distance we expect no significant improvement with respect to the
previous case.

Notice that these considerations can be drawn by looking at the
equiprobability curves for neutrinos and antineutrinos, only. We will
see in the following section how the theoretical expectation is indeed
reproduced in the ``experimental data''.

\section{Number of ``wrong-sign'' muons in the detector}
\label{sec:emuENE}

The experimental information is not the transition probability
$P^\pm_{e\mu}$ but the number of muons with charge opposite to that of
the muons circulating in the storage ring, that in the following will
be often called ``golden'' muons.  The events are then grouped in bins
of energy, with the size of the energy bin depending on the energy
resolution $\Delta E$ of the considered detector.  In general, 
\be
N^i_{\mu^\mp} (\bar \theta_{13}, \bar \delta) = \int_{E_i}^{E_i +
\Delta E} dE \, \sigma_{\nu_\mu (\bar \nu_\mu)} (E) \, P^\pm_{e\mu}
(E, \bar \theta_{13}, \bar \delta) \, \frac{d \Phi_{\nu_e (\bar \nu_e)
} (E)}{d E}
\label{eq:evento}
\ee 
is the number of wrong-sign muons in the i-th energy bin for the
input pair ($\bar \theta_{13}, \bar \delta$); $E$ is the neutrino
(antineutrino) energy\footnote{The neutrino energy can be reconstructed 
if the considered detector has a hadronic calorimeter capable to measure 
the energy of the hadronic shower ($E_{hadr}$) in the $\nu N$ CC
interactions with good precision.}.
The charged current neutrino and antineutrino
interaction rates can be computed using the approximate expressions
for the neutrino-nucleon cross sections on an isoscalar target, \be
\sigma_{\nu N} \sim 0.67 \times 10^{-42} \times \frac{E_\nu}{GeV}
\times m^2 \, ; \qquad \sigma_{\bar \nu N} \sim 0.34 \times 10^{-42}
\times \frac{E_\nu}{GeV} \times m^2 \ .
\label{isoscalar}
\ee In the laboratory frame the neutrino fluxes, boosted along the
muon momentum vector, are given by: \bea \frac{ d^2 \Phi_{\nu_e,
\bar\nu_e} }{ dy d\Omega} & = & \frac{ 24 n_\mu }{ \pi L^2 m_\mu^6 }
\,\, \bar E_\mu^4 y^2 \, (1 - \beta \cos \varphi) \,\, \left \{ \left
[ m_\mu^2 - 2 \bar E_\mu^2 y \, (1 - \beta \cos \varphi) \right ]
\right . \nn \\ & & \left. \mp \, {\cal P}_\mu \left [ m_\mu^2 - 2
\bar E_\mu^2 y \, (1 - \beta \cos \varphi) \right ] \right \} \, .
\label{fluxes}
\eea Here, $\beta = \sqrt{1-(m_\mu/\bar E_\mu)^2}$, $\bar E_\mu$ is
the parent muon energy, $y = E_\nu/\bar E_\mu$, $n_\mu$ is the number
of useful muons per year obtained from the storage ring and $\varphi$
is the angle between the beam axis and the direction pointing towards
the detector. In what follows, the fluxes have been integrated in the
forward direction with an angular divergence (taken to be constant)
$\delta \varphi \sim 0.1$ mr. The effects of the beam divergence and
the QED one-loop radiative corrections to the neutrino fluxes have
been properly taken into account in \cite{Broncano:2002hs}. The
overall correction to the neutrino flux has been shown to be of ${\cal
O} (0.1 \%)$.

In the same approximations as for eq.~(\ref{eq:spagnoli}), we get for the number of events
per bin: 
\be
\label{eq:numeroeventi}
\left \{
\begin{array}{lll}
N^i_{\mu^-} &=& I^i_{X_+} \sin^2 ( 2 \theta_{13}) + 
                \left [ I^i_{Y^c_+} \cos \delta + I^i_{Y^s_+} \sin \delta \right ]
                \cos \theta_{13} \sin (2 \theta_{13}) +
                I^i_Z \, , \\
&& \\
N^i_{\mu^+} &=& I^i_{X_-} \sin^2 ( 2 \theta_{13}) + 
                \left [ I^i_{Y^c_-} \cos \delta - I^i_{Y^s_-} \sin \delta \right ]
                \cos \theta_{13} \sin (2 \theta_{13}) +
                I^i_Z \, , 
\end{array}
\right .
\ee
where we introduced a short-form notation for the following integrals:
\be
\left \{
\begin{array}{lll}
I^i_{X_\pm} &=& \int_{E_i}^{E_i + \Delta E} dE \, \sigma_{\nu_\mu
                (\bar \nu_\mu)} \frac{d \Phi_{\nu_e (\bar \nu_e) }
                (E)}{d E} X_\pm (E) \, , \\ && \\ I^i_{Y^c_\pm} &=&
                \int_{E_i}^{E_i + \Delta E} dE \, \sigma_{\nu_\mu
                (\bar \nu_\mu)} \frac{d \Phi_{\nu_e (\bar \nu_e) }
                (E)}{d E} Y_\pm (E) \cos \left ( \frac{\Delta_{atm}
                L}{2}\right ) \, , \\ && \\ I^i_{Y^s_\pm} &=&
                \int_{E_i}^{E_i + \Delta E} dE \, \sigma_{\nu_\mu
                (\bar \nu_\mu)} \frac{d \Phi_{\nu_e (\bar \nu_e) }
                (E)}{d E} Y_\pm (E) \sin \left ( \frac{\Delta_{atm}
                L}{2}\right ) \, , \\ && \\ I^i_Z &=& \int_{E_i}^{E_i
                + \Delta E} dE \, \sigma_{\nu_\mu (\bar \nu_\mu)}
                \frac{d \Phi_{\nu_e (\bar \nu_e) } (E)}{d E} Z (E) \,
                .
\end{array}
\right .
\label{eq:integrals}
\ee

For a fixed energy bin and fixed input parameters ($\bar \theta_{13},
\bar \delta$), we can draw a continuous curve of equal number of
events in the ($\Delta \theta, \delta$) plane, \be N^i_{\mu^\pm}
(\theta_{13}, \delta) = N^i_{\mu^\pm} (\bar \theta_{13}, \bar
\delta)\, , \ee as it was the case for the transition probability,
eq.~(\ref{eq:equi0}).

\begin{figure}[h!]
\begin{center}
\begin{tabular}{ccc}
\hspace{-1cm} \epsfxsize5cm\epsffile{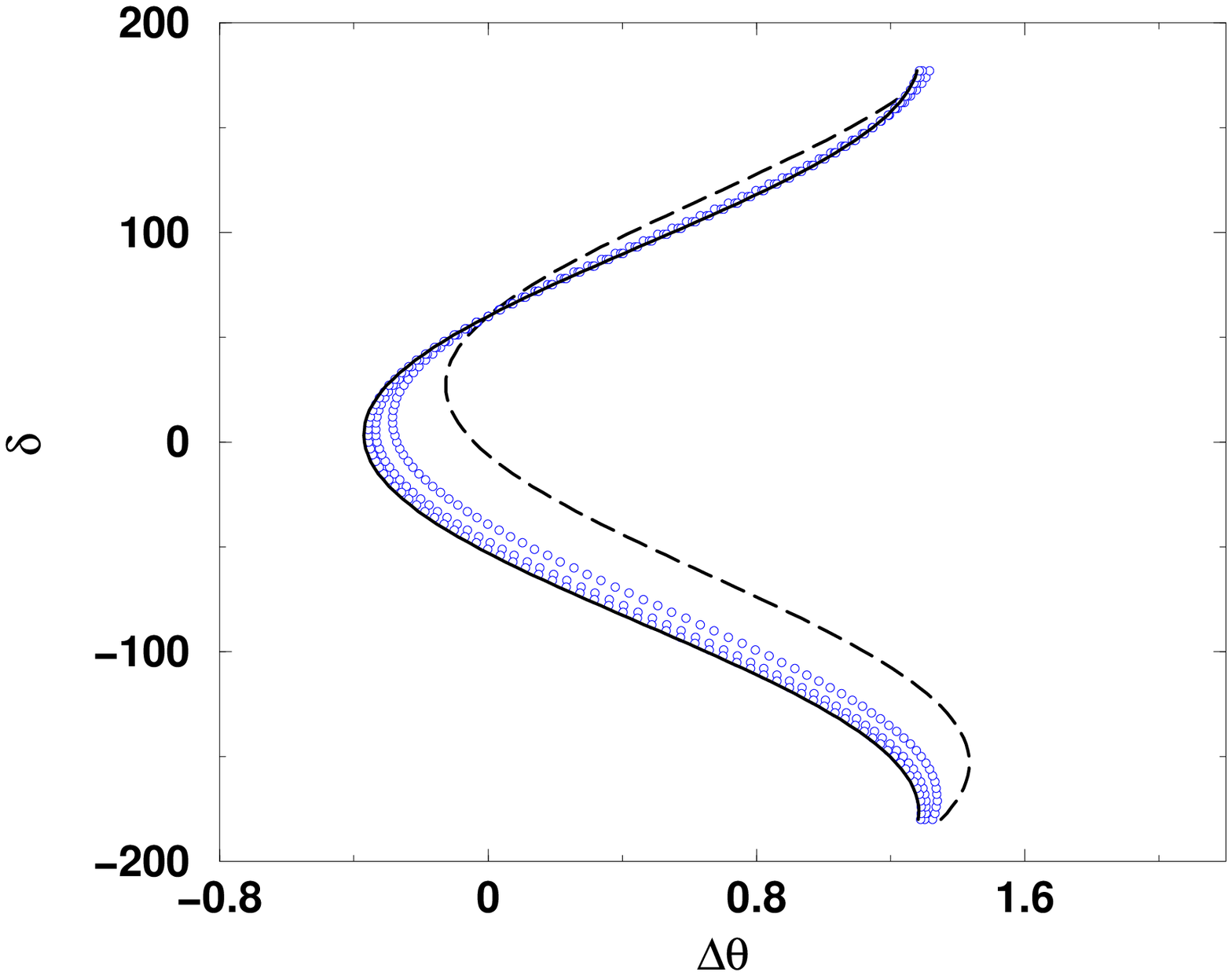} &
              \epsfxsize5cm\epsffile{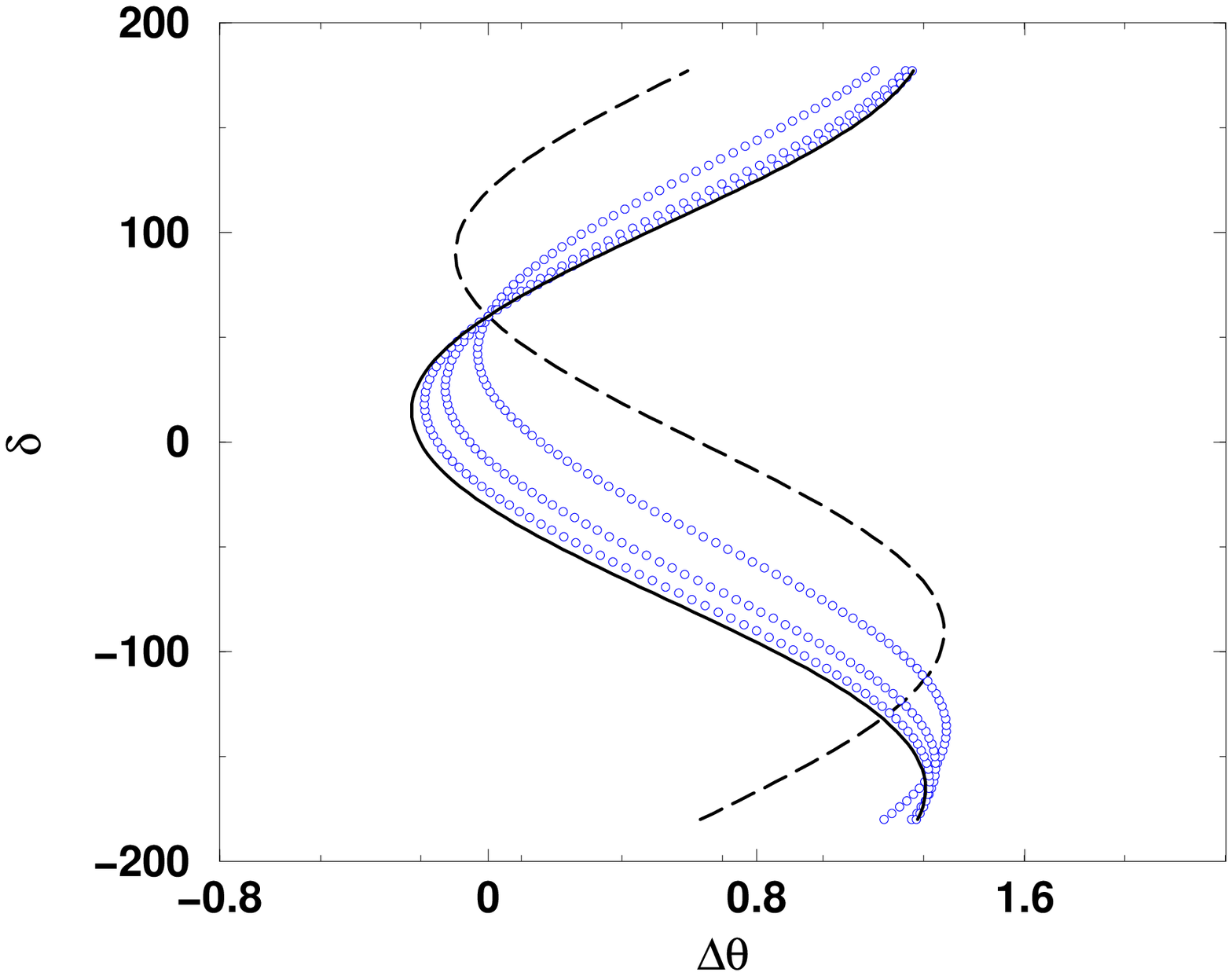} &
              \epsfxsize5cm\epsffile{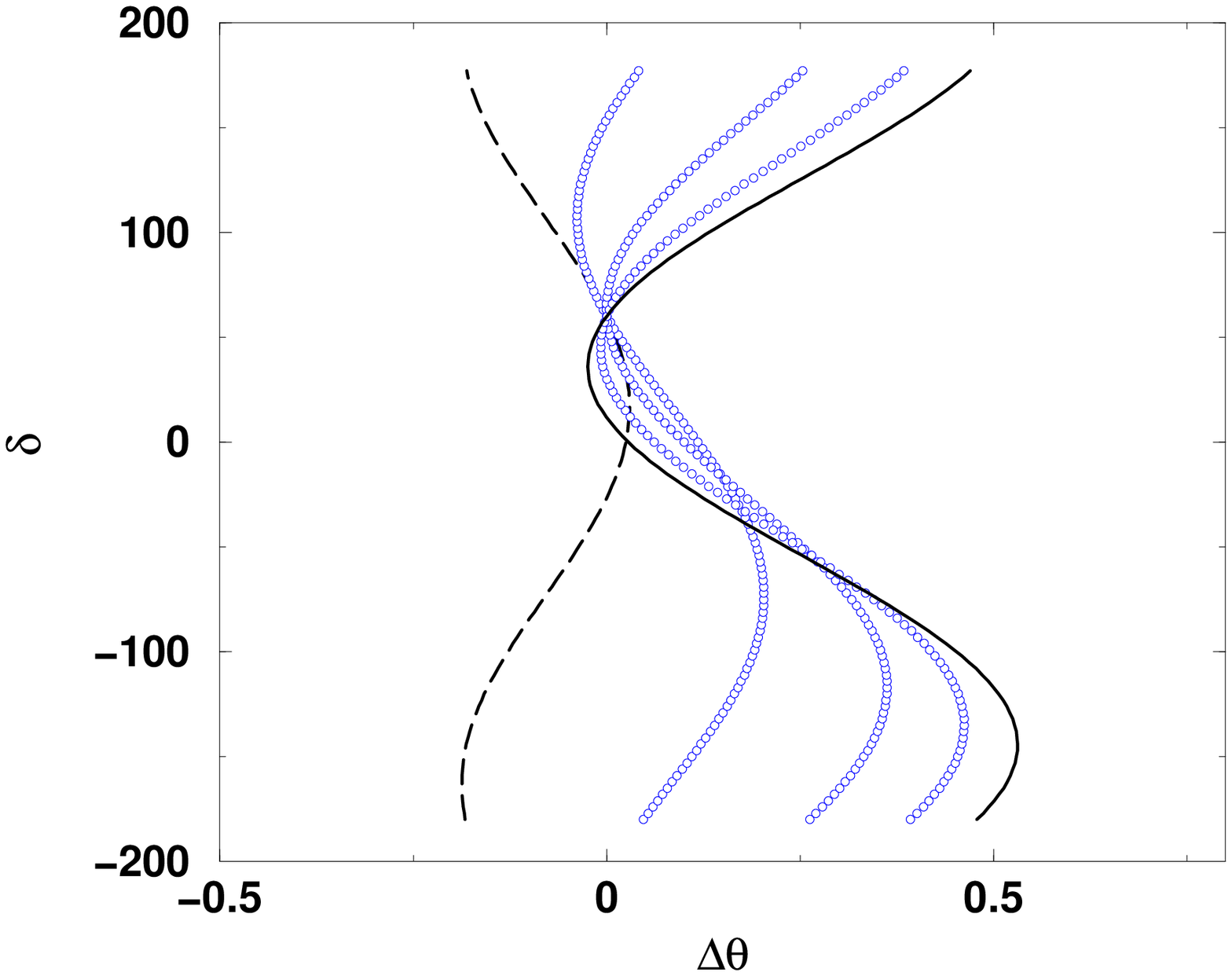}
\end{tabular}
\caption{\it Equal-number-of-events (ENE) curves in the ($\Delta \theta, \delta$) plane for 
neutrinos, for $\bar \theta_{13} = 5^\circ, \bar \delta = 60^\circ$, 
$E_\nu \in [5, 50] $ GeV and $L = 732,\, 3000$ and $7332$ Km.
The dashed line represents the first energy bin, $E_\nu \in [0, 10]$ GeV, 
the solid line the last energy bin, $E_\nu \in [40, 50]$ GeV; 
the dotted lines lie in between these two.}
\label{fig:equievents}
\end{center}
\end{figure}

\noindent We therefore get an implicit equation in $\delta$, \be F(
\delta ) = G ( \theta_{13}, \bar \theta_{13}, \bar \delta ) \, , \ee
where \be \left \{ \begin{array}{lll} F ( \delta ) &=& \cos \delta \pm
\left ( \frac{I^i_{Y^s_\pm} }{I^i_{Y^c_\pm}} \right ) \sin \delta \, ,
\\ G ( \theta_{13}, \bar \theta_{13}, \bar \delta) &=& \left (
\frac{I^i_{X_\pm} }{I^i_{Y^c_\pm}} \right ) f (\theta_{13}, \bar
\theta_{13} ) + F (\bar \delta ) g( \theta_{13}, \bar \theta_{13} )
\end{array} \right .  \ee and $f(\theta_{13}, \bar \theta_{13})$ and
$g(\theta_{13}, \bar \theta_{13})$ are the $\theta$-dependent
functions introduced in eq.~(\ref{eq:functions}).  Solving for
$\delta$, \be \delta = F^{-1} [ G(\theta_{13}, \bar\theta_{13}, \bar
\delta ) ] \ee we get equal-number-of-events curves (ENE) in the
($\Delta \theta, \delta$) plane, see Fig.~\ref{fig:equievents}.

The problem arises in the reconstruction of the physical parameters
from a data set consisting of some given number of events per bin, for
a given number of bins (depending on the specific detector energy
resolution).  As it was the case for the equiprobability curves in the
previous section, all the ENE curves intersect in the physical point
($\Delta \theta = 0, \delta = \bar \delta$) and any given couple of
curves intersect in a second point in the same region as in
Fig.~\ref{fig:equiprob}. As it can be seen in
Fig.~\ref{fig:equievents}, the second intersection differs when
considering different couples of curves, but lies always in a
restricted area of the ($\Delta \theta, \delta$) plane, the specific
location of this region depending on the input parameters ($\bar
\theta_{13}, \bar \delta$), see eqs.~(\ref{eq:deltapapprx}) and
(\ref{eq:deltamapprx}).  The $\chi^2$ analysis of the data will
therefore identify two allowed regions: the ``physical'' one (around
the physical value, $\bar \theta_{13}, \bar \delta$) and the ``clone''
solution, spanning all the area where a second intersection between
any two ENE curves occurs.  This is the source of the ambiguity
pointed out in \cite{Burguet-Castell:2001ez}.

In the remaining of this section, we apply the analysis in energy bins
of \cite{Cervera:2000kp,Burguet-Castell:2001ez}. Let $N_{i,p}^\lambda$
be the total number of wrong-sign muons detected when the factory is
run in polarity $p=\mu^+,\mu^-$, grouped in energy bins specified by
the index $i$, and three possible distances, $\lambda =$ 1, 2, 3
(corresponding to $L = 732$ Km, $L = 3000$ Km and $L = 7332$ Km,
respectively).  In order to simulate a typical experimental situation
we generate a set of ``data'' $n_{i,p}^\lambda$ as follows: for a
given value of the oscillation parameters, the expected number of
events, $N_{i,p}^\lambda$, is computed; taking into account
backgrounds and detection efficiencies per bin, $b_{i,p}^\lambda$ and
$\epsilon_{i,p}^\lambda$, we then perform a gaussian (or poissonian
for $N_{i,p}^\lambda \le 10$ events) smearing to mimic the statistical
uncertainty: \be
\label{eq:smearing}
n_{i,p}^\lambda = {\rm Smear} (N_{i,p}^\lambda \epsilon_{i,p}^\lambda
+ b_{i,p}^\lambda) \,.  \ee Finally, ``data'' are fitted to the
theoretical expectation as a function of the neutrino parameters under
study, using a gaussian $\chi^2$ minimization: \be
\label{eq:chi2eventi}
\chi_\lambda^2 = \sum_p \sum_i 
\left(\frac{ n^\lambda_{i,p} \, - \, 
N^{\lambda}_{i,p}}{\delta n^\lambda_{i,p}}\right)^2 \, ,
\label{chi2}
\ee where $\delta n^\lambda_{i,p}$ is the statistical error for
$n^\lambda_{i,p}$ (errors on background and efficiencies are
neglected) or a poissonian $\chi^2$ minimization: \be
\label{eq:chi2poisson}
\chi_\lambda^2 = - 2 \sum_p \sum_i \left [ \left ( n^\lambda_{i,p} -
N^\lambda_{i,p} \right ) - n^\lambda_{i,p} \log \left (
\frac{n^\lambda_{i,p}}{N^\lambda_{i,p} } \right ) \right ] \ee
whenever events are Poisson-distributed around the theoretical values
(see \cite{Huber:2002mx} and refs. therein).  We verified that the
fitting of theoretical numbers to the smeared (``experimental'') ones
is able to reproduce the values of the input parameters (the best fit
always lies within a restricted region around $\bar \theta_{13}, \bar
\delta$).

The following ``reference set-up'' has been considered: neutrino beams
resulting from the decay of $ 2 \times 10^{20} \mu^+$'s and $\mu^-$'s
per year in a straight section of an $\bar E_\mu = 50$ GeV muon
accumulator. An experiment with a realistic 40 Kton detector of
magnetized iron and five years of data taking for each polarity is
envisaged.  Detailed estimates of the corresponding expected
backgrounds and efficiencies have been included in the analysis,
following \cite{Cervera:2000vy}.  Notice that this set-up is exactly
the same of \cite{Cervera:2000kp,Burguet-Castell:2001ez}.

\begin{figure}[h!]
\begin{center}
\begin{tabular}{c}
\hspace{-0.3cm}
\epsfxsize12.8cm\epsffile{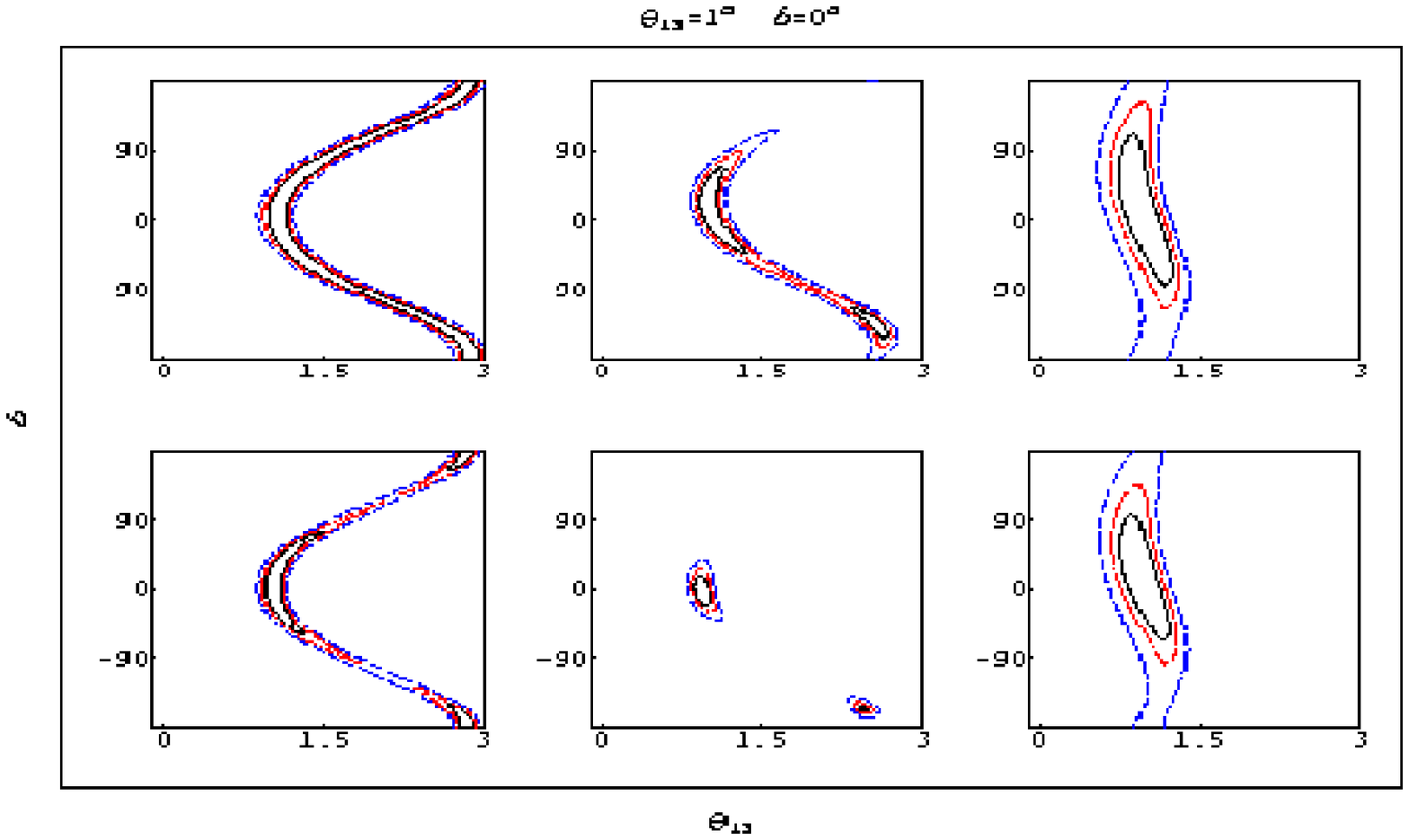} \\
\epsfxsize12.8cm\epsffile{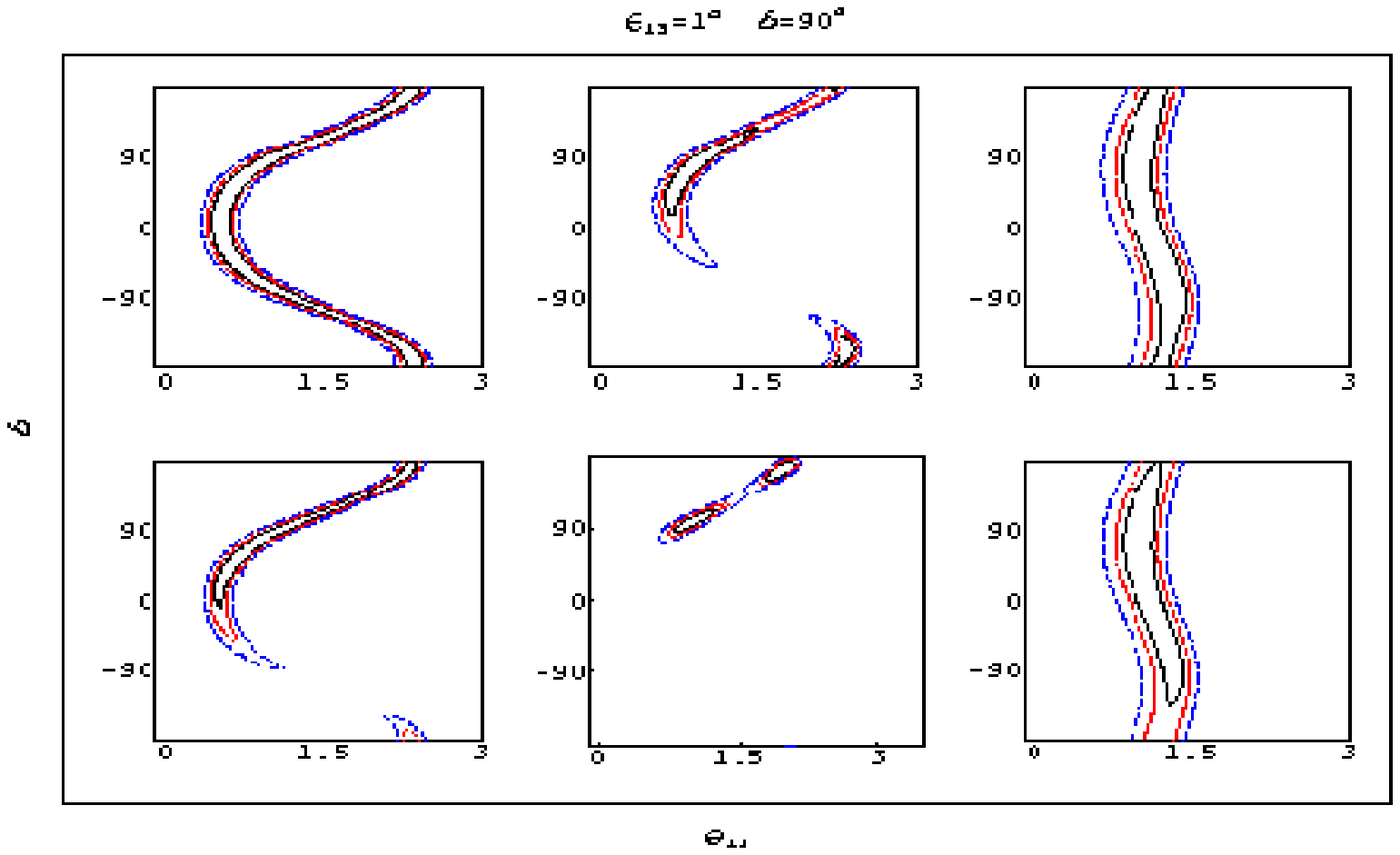}
\end{tabular}
\vspace{0.5cm}
\caption{\it 68.5, 90 and 99 \% C.L. contours resulting from a
$\chi^2$ fit of $\theta_{13}$ and $\delta$, for $\bar \theta_{13} =
1^\circ$ and $\bar \delta = 0^\circ$ (up) and $\bar \delta = 90^\circ$
(down).  For each considered input parameters couple, the two rows
represent: from left to right, $L = 732, 3000$ and $7332$ Km; the upper
row is $N^i_{\mu^+}, i = 1, \dots, 5$; the lower row is $N^i_{\mu^+}$
and $N^i_{\mu^-}, i = 1, \dots, 5$.}
\label{fig:th=1}
\end{center}
\end{figure}

\begin{figure}[h!]
\begin{center}
\begin{tabular}{c}
\epsfxsize12.5cm\epsffile{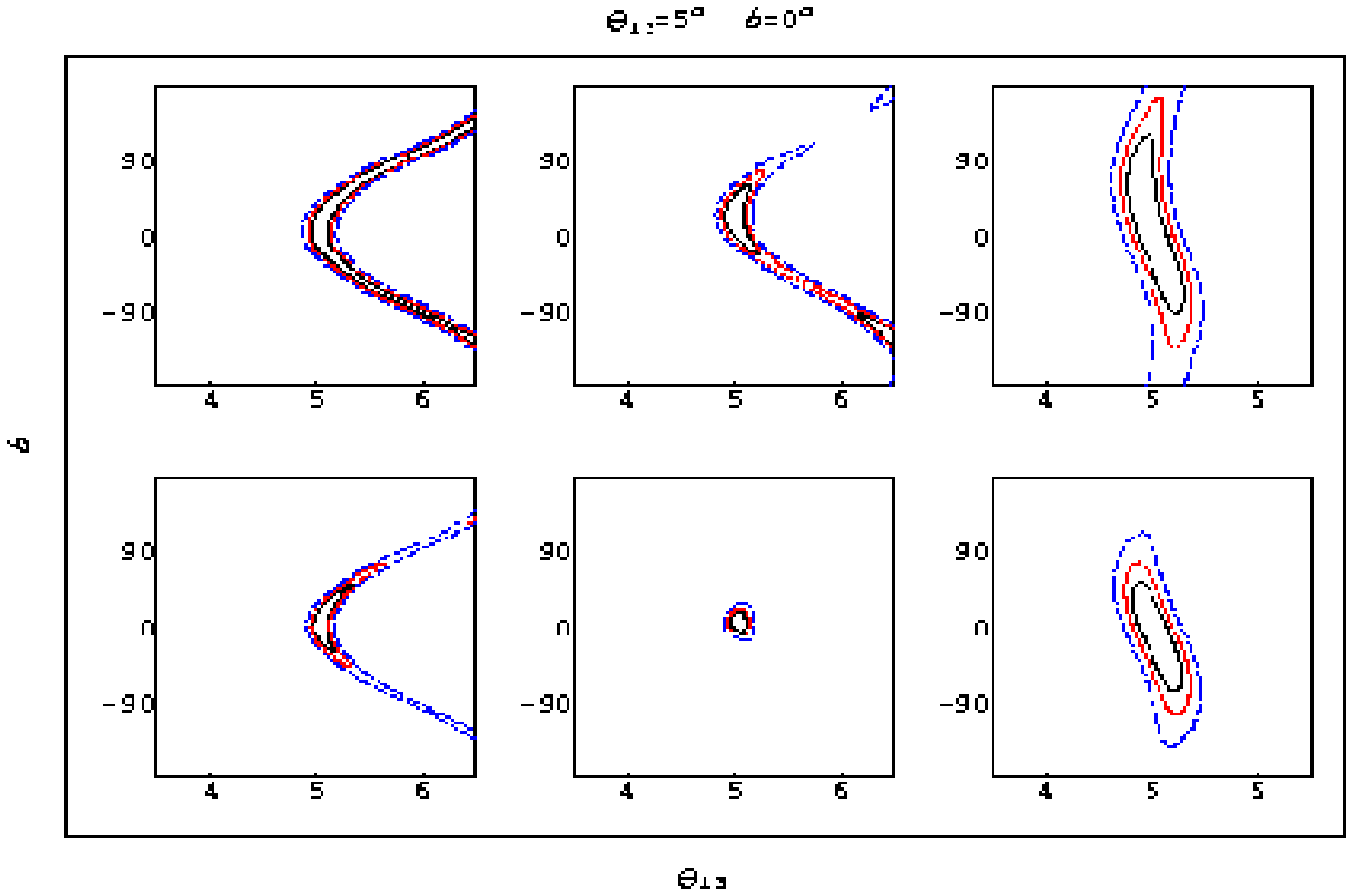} \\
\epsfxsize12.5cm\epsffile{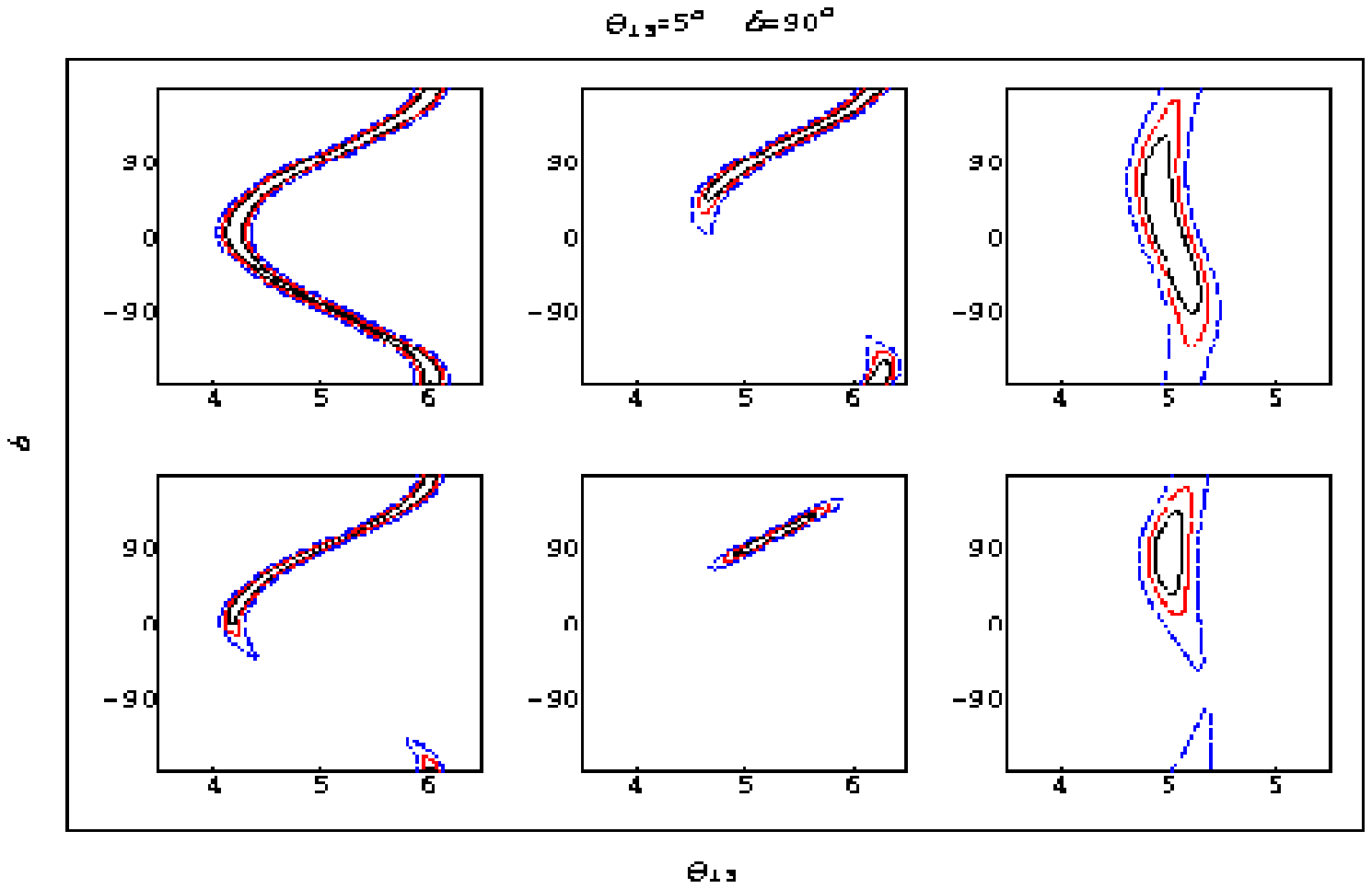}
\end{tabular}
\vspace{0.5cm}
\caption{\it 68.5, 90 and 99 \% C.L. contours resulting from a
$\chi^2$ fit of $\theta_{13}$ and $\delta$, for $\bar \theta_{13} =
5^\circ$ and $\bar \delta = 0^\circ$ (up) and $\bar \delta = 90^\circ$
(down).  For each considered input parameters couple, the two rows
represent: from left to right, $L = 732, 3000$ and $7332$ Km; the upper
row is $N^i_{\mu^+}, i = 1, \dots, 5$; the lower row is $N^i_{\mu^+}$
and $N^i_{\mu^-}, i = 1, \dots, 5$.}
\label{fig:th=5}
\end{center}
\end{figure}

In the first row of Figs. \ref{fig:th=1}-\ref{fig:th=5}, we present
the results of the fit to five bins of data for decaying muons of one
single polarity, $\mu^-$.  The energy resolution of the detector is
$\Delta E_\nu = 10$ GeV.  In all cases we observe the pattern depicted
in the previous section, with a good determination of $\bar
\theta_{13}$ and an extremely poor determination of $\bar \delta$.  In
the second row we fit to five bins of data for decaying muons of both
polarities. The results follow again the theoretical analysis of this
and of the previous section and are in perfect agreement with what
presented in \cite{Burguet-Castell:2001ez}.  In particular, notice how
at the intermediate distance it is possible now to reconstruct
$\delta$ with an error of tens of degrees in two separate regions of
the parameter space.

\newpage

\newpage

\newpage

\section{The $\nu_e \to \nu_\tau$ equiprobability and ENE curves}
\label{sec:etau}

We present in this section the possibility to use a different channel,
namely the $\nu_e \to \nu_\tau$ oscillation probability, to improve
the reconstruction of the physical parameters ($\bar \theta_{13}, \bar
\delta$) in combination with the results for the $\nu_e \to \nu_\mu$
transition described in \cite{Cervera:2000kp,Burguet-Castell:2001ez}
and in the previous section.

The $\nu_e \to \nu_\tau$ oscillation probability at second order in
perturbation theory in $\theta_{13}$, $\Delta_\odot/\Delta_{atm}$,
$\Delta_\odot/A$ and $\Delta_\odot L$ is: \be
\label{eq:etau}
P^\pm_{e \tau} (\bar \theta_{13}, \bar \delta) = 
X^\tau_\pm \sin^2 (2 \bar \theta_{13}) -
Y_\pm \cos ( \bar \theta_{13} ) \sin (2 \bar \theta_{13} )
      \cos \left ( \pm \bar \delta - \frac{\Delta_{atm} L }{2} \right ) 
+ Z^\tau \, ,
\ee
where $\pm$ refers to neutrinos and antineutrinos, respectively, and
\be
\label{eq:etaucoeff}
\left \{
\begin{array}{lll}
X^\tau_\pm &=& \cos^2 (\theta_{23} ) \left ( \frac{\Delta_{atm} }{ B_\mp } \right )^2
                                       \sin^2 \left ( \frac{ B_\mp L}{ 2 } \right ) \ , \\
\nn \\
Y_\pm &=& \sin ( 2 \theta_{12} ) \sin ( 2 \theta_{23} )
                                     \left ( \frac{\Delta_\odot }{ A } \right )
                                     \left ( \frac{\Delta_{atm} }{ B_\mp } \right )
                                         \sin \left ( \frac{A L }{ 2 } \right )
                                         \sin \left ( \frac{ B_\mp L }{ 2 } \right ) \ , \\
\nn \\
Z^\tau &=& \sin^2 (\theta_{23} ) \sin^2 (2 \theta_{12})
\left ( \frac{\Delta_\odot }{ A } \right )^2
\sin^2 \left ( \frac{A L }{ 2 } \right ) \ ,
\end{array}
\right .  \ee with $Z^\tau = Z^\tau_+ = Z^\tau_-$.  Notice that
$X^\tau_\pm$ and $Z^\tau$ differs from the corresponding coefficients
for the $\nu_e \to \nu_\mu$ transition for the $\cos \theta_{23}
\leftrightarrow \sin \theta_{23}$ exchange, only. The $Y_\pm$ term is
identical for the two channels, but it appears with an opposite
sign. This sign difference in the $Y$-term is crucial, as it
determines a different shape in the $(\Delta \theta, \delta)$ plane
for the two sets of equiprobability curves.

\begin{figure}[h!]
\begin{center}
\begin{tabular}{cc}
\hspace{-1cm} \epsfxsize7cm\epsffile{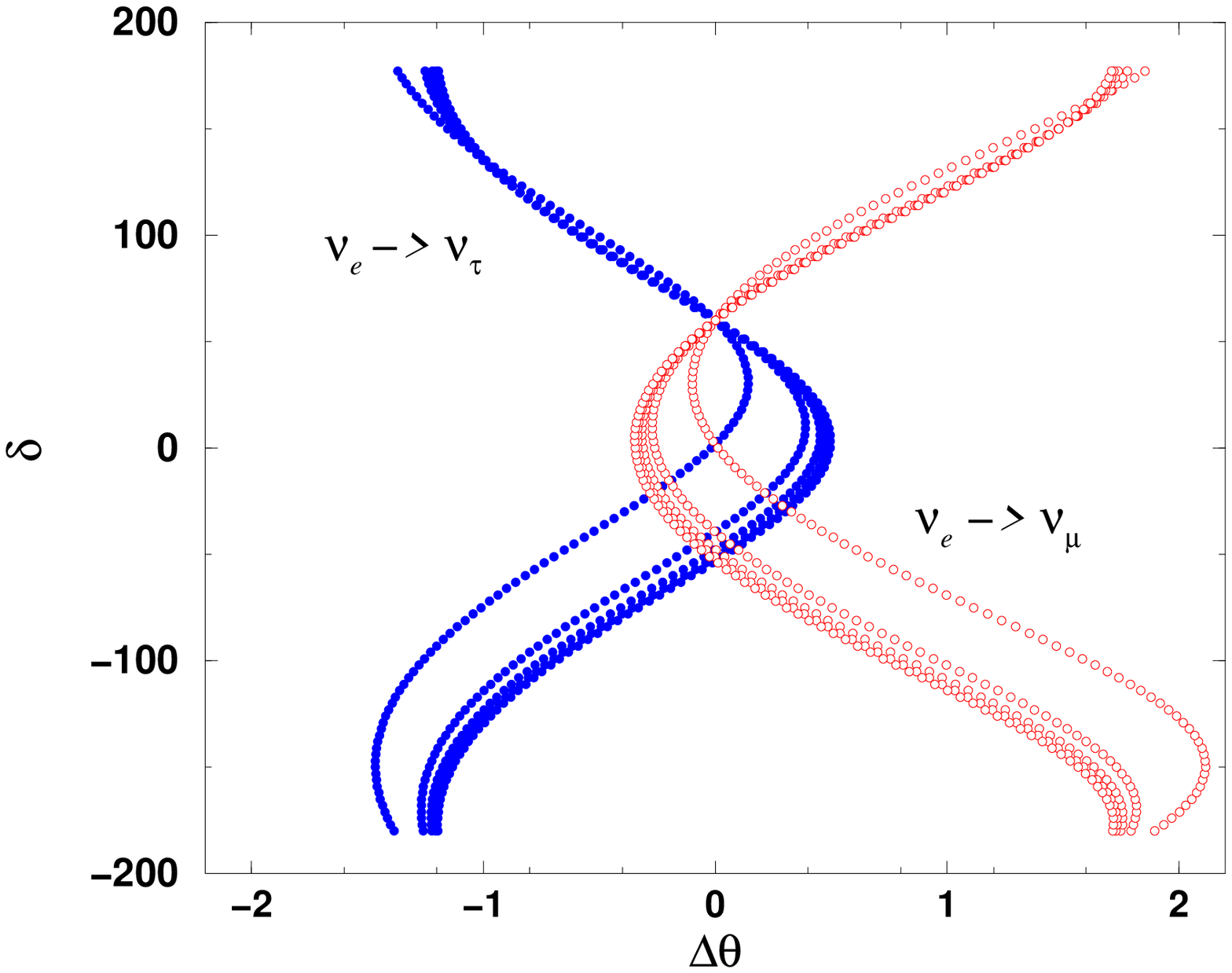} &
              \epsfxsize7cm\epsffile{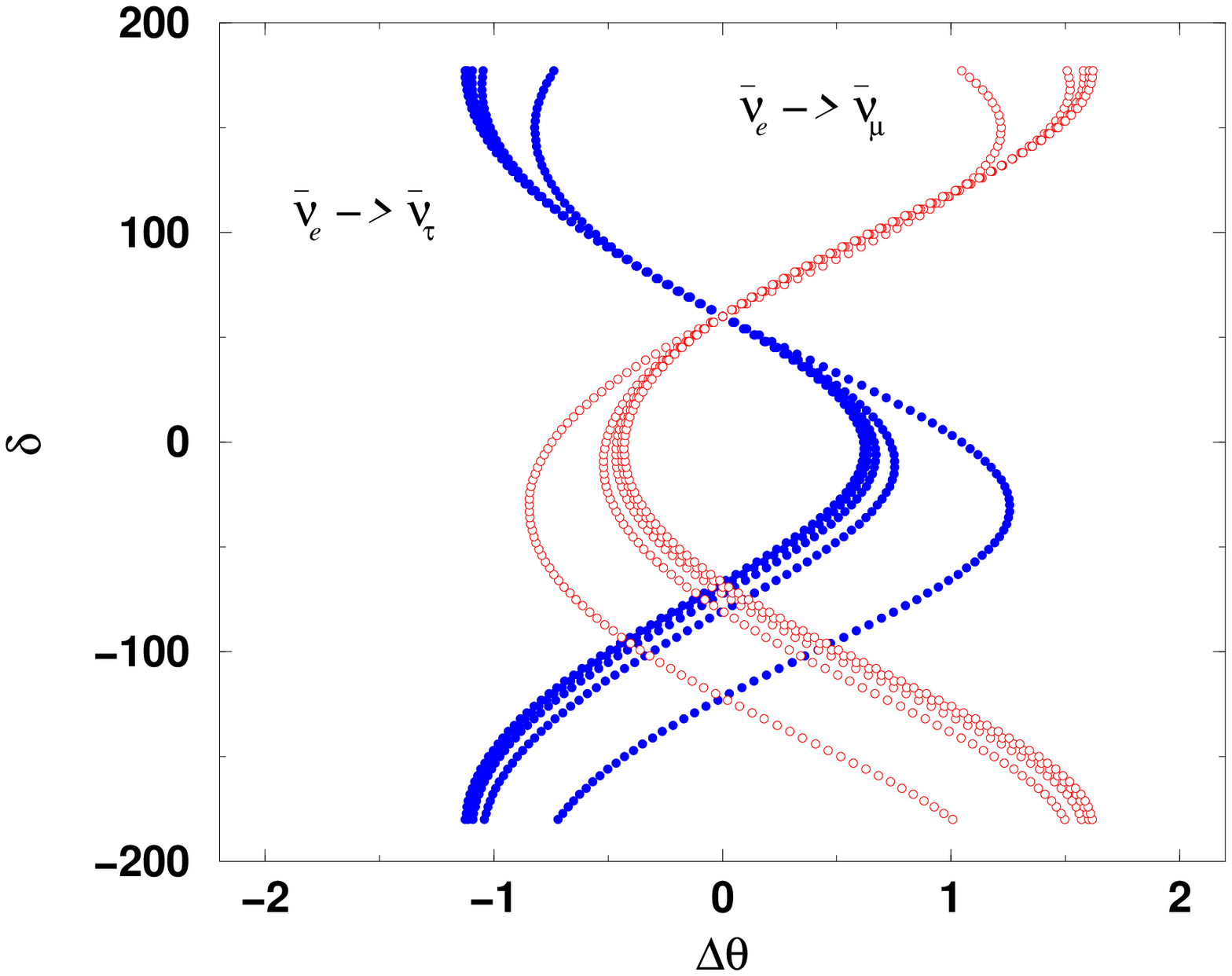}
\end{tabular}
\caption{\it Equiprobability curves in the ($\Delta \theta, \delta$) plane, 
for $\bar \theta_{13} = 5^\circ, \bar \delta = 60^\circ$, $E_\nu \in [5, 50] $ GeV
and $L = 732$ Km for the $\nu_e \to \nu_\mu$ and $\nu_e \to \nu_\tau$ oscillation
(neutrinos on the left, antineutrinos on the right).}
\label{fig:equiprobetau}
\end{center}
\end{figure}

In Fig.~\ref{fig:equiprobetau}, we superimposed the equiprobability
curves for the $\nu_e \to \nu_\tau$ and $\nu_e \to \nu_\mu$
oscillations at a fixed distance, $L = 732$ Km, with input parameters
$\bar \theta_{13} = 5^\circ$ and $\bar \delta = 60^\circ$, for
different values of the energy, $E_\nu \in [5, 50]$ GeV. The effect of
the different sign in front of the $Y$-term in
eqs.~(\ref{eq:spagnoli}) and (\ref{eq:etau}) can be seen in the
opposite shape in the ($\theta_{13}, \delta$) plane of the $\nu_e \to
\nu_\tau$ curves with respect to the $\nu_e \to \nu_\mu$ ones. Notice
that all the curves of both families met in the ``physical'' point,
$\theta_{13} = \bar \theta_{13}$, $\delta = \bar \delta$, and that now
three would-be ``clone'' regions (i.e., the spread regions where the
intersections of any given couple of equiprobability curves lie) can
be seen.

As a final comment we signal that, if $\theta_{13}$ is not extremely small
(in such a way that the $X_\pm$ and $X^\tau_\pm$ terms dominate over the $Y_\pm$ terms
in eqs.~(\ref{eq:spagnoli}) and (\ref{eq:etau})), the combined measurement of $\nu_e \to \nu_\mu$
and $\nu_e \to \nu_\tau$ transitions could in principle solve 
the $[\theta_{23}, \pi/2 - \theta_{23}]$ ambiguity.

To follow the line of reasoning adopted for the $\nu_e \to \nu_\mu$
channel, we should now discuss how the $\nu_e \to \nu_\tau$ channel
can be used in a realistic experiment.  A number of modifications with
respect to the case of the ``golden'' channel should be taken into
account.

First, the approximate expressions for the neutrino-nucleon cross
section on an isoscalar target, eq.~(\ref{isoscalar}), are no longer
appropriate in the case of a $\nu_\tau$ CC interaction inside the
detector. In this case we used the reported values for the $\nu_\tau
N$ cross-section \cite{httpchorus} that have been applied in the
CHORUS and OPERA experiment to compute the expected number of CC
$\tau$ events.  The considered cross-section includes $\tau$ mass
effects in the DIS region following \cite{Albright:1975ts} (see
App.~\ref{app:form} for details), as well as the elastic and
quasi-elastic contributions to the cross section.

Second, the $\tau$ will decay in flight into a muon of the same charge
and two neutrinos, with a branching ratio $BR (\tau \to \mu) \simeq
0.17$, \cite{PDG}. This ``silver'' wrong-sign muon is the experimental
signal we are looking for, to be identified and to be separated from
the ``golden'' wrong-sign muons originated from $\nu_\mu$ CC
interactions\footnote{We adopt the nick-name of ``silver'' muon events
due to the lesser statistical significance with respect to ``golden''
ones.}.  The first tool to distinguish the two sets of wrong-sign
muons is their different energy distribution (see App.~\ref{app:form}
for details on the differential decay rate).  It has been shown in
\cite{Cervera:2000vy} that in the magnetized iron detector considered
in the previous section, muons from $\tau$ decay cannot be
distinguished from the main background represented by muons from
charmed mesons decay by means of kinematical cuts.  
In order to take advantage of this channel, we
should therefore use a different kind of detector: for this reason we
concentrate in the remaining of the paper on a lead-emulsion detector,
where the observation of the $\tau$ decay vertex allows to
distinguish ``golden'' and ``silver'' wrong-sign muons, and the
latter from the charmed mesons decay background.
We must mention that the $\nu_e \to \nu_\tau$ oscillations were previously
considered in \cite{Bueno:2000fg} for a liquid argon detector, using
kinematical cuts to identify $\tau$'s. It could be of interest to explore 
further the possibility of using ``silver'' muon events in such a detector
to reduce or eliminate the ($\theta_{13},\delta$) ambiguity.

In what follows, we consider an OPERA-like detector with a
mass of 2 Kton and spectrometers capable of muon charge identification
(see the OPERA proposal for details, \cite{Guler:2000bd}) located at
$L = 732$ Km down the neutrino source (obviously, in the back of our
mind we are thinking of the CNGS set-up).  The results for both the
``golden'' and the ``silver'' channel at the near emulsion detector
will be combined with results for the ``golden'' channel obtained with
the magnetized iron detector located at the optimal distance for the
measurement of leptonic CP violation, $L = 3000$ Km.

In this paper, we will first restrict ourselves to an ideal OPERA-like
detector with perfect efficiency and no background. Afterwards, we
take into account the realistic estimates of the energy-dependent
reconstruction efficiency and of the most relevant
backgrounds\footnote{ A dedicated careful analysis of the ``silver''
muons reconstruction efficiency and of the background as a function of
the neutrino energy for this specific detector is currently under
progress, \cite{Autieroetal}. A key issue is the maximum affordable
amount of charge discrimination, both for the emulsion and the
magnetized iron detector.  Estimates can be found in
\cite{Cervera:2000vy} for the magnetized iron detector and in Fig.~86
of \cite{Guler:2000bd} for the emulsion detector.}.  Eventually, we
will consider how an increase in the detector mass or an improvement
on the signal/noise ratio affects our results.

\begin{figure}
\begin{center}
\epsfxsize10cm\epsffile{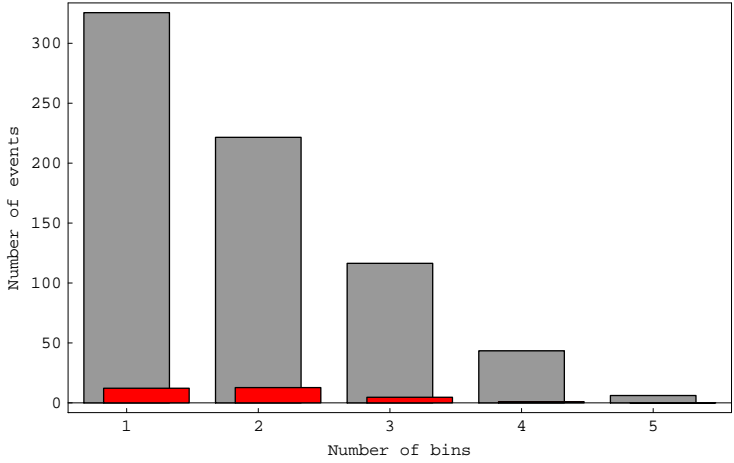} 
\caption{\it Comparison of the number of ``golden'' muons (lightest
bars) and ``silver'' muons (darkest bars), for a parent muon energy
$\bar E_\mu = 50$ GeV and input parameters $\bar \theta_{13} =
5^\circ$, $\bar \delta = 60^\circ$ and a near detector ($L = 732$ Km)
with a mass of 2 Kton and perfect reconstruction efficiency for both
channels.}
\label{fig:mudistr}
\end{center}
\end{figure}

Schematically, starting from a positive charged muon in the storage ring, 
``silver'' muons are obtained by the following chain: 
\bea
\mu^+ \to \left \{ 
\begin{array}{ccccccc}
e^+ & & & & & &\\
\bar \nu_\mu & & & & & & \\
\nu_e & \to & \nu_\tau & \to & \tau^- & \to & \mu^-
\end{array}
\right . \nn \\
\nn
\eea
whereas ``golden'' muons come from: 
\bea
\mu^+ \to \left \{ 
\begin{array}{ccccc}
e^+ & & & & \\
\bar \nu_\mu & & & & \\
\nu_e & \to & \nu_\mu & \to & \mu^- 
\end{array}
\right . \nn \\
\nn
\eea
If we group the events in bins of the final muon energy $E_\mu$, with the size of the 
energy bin depending on the energy resolution $\Delta E_\mu$ of the considered detector, 
the number of ``golden'' muons in the i-th energy bin for the input pair 
($\bar \theta_{13}, \bar \delta$) and for a parent muon energy $\bar E_\mu$ is:
\be
N^g_{\mu^\mp} (\bar \theta_{13}, \bar \delta)
= \left \{ \frac{d \sigma_{\nu_\mu (\bar \nu_\mu)} (E_\mu, E_\nu) }{d E_\mu}
                           \, \otimes \,
                 P^\pm_{e\mu} (E_\nu, \bar \theta_{13}, \bar \delta)           
                           \, \otimes \,
                 \frac{d \Phi_{\nu_e (\bar \nu_e) } (E_\nu, \bar E_\mu)}{d E_\nu} 
                      \right \}_{E_i}^{E_i + \Delta E_\mu} 
\label{eq:convogolden}
\ee
(remember that $P^\pm$ is the oscillation probability for neutrinos and antineutrinos, 
respectively, see Sect.~\ref{sec:emuprob}), whereas the number of ``silver'' muons 
in the i-th energy bin is:
\bea
N^s_{\mu^\mp} (\bar \theta_{13}, \bar \delta)
&=& BR(\tau \to \mu) \, 
   \left \{ \left [ 
      \frac{d N_{\mu^\mp} (E_\mu, E_\tau) }{d E_\mu}                    
                           \, \otimes \,
      \frac{d \sigma_{\nu_\tau (\bar \nu_\tau)} (E_\tau, E_\nu) }{d E_\tau}
           \right ] \right .\nn \\
& & \left .  \qquad             \, \otimes \,
                 P^\pm_{e\tau} (E_\nu, \bar \theta_{13}, \bar \delta)           
                           \, \otimes \,
                 \frac{d \Phi_{\nu_e (\bar \nu_e) } (E_\nu, \bar E_\mu)}{d E_\nu} 
                      \right \}_{E_i}^{E_i + \Delta E_\mu} \, .
\label{eq:convosilver}
\eea In both equations, $\otimes$ stands for a convolution integral on
the intermediate energy: for example, 
\bea
BR(\tau \to \mu) \left [
\frac{d N_{\mu^\mp} (E_\mu, E_\tau) }{d E_\mu} \, \otimes \, \frac{d
\sigma_{\nu_\tau (\bar \nu_\tau)} (E_\tau, E_\nu) }{d E_\tau} \right ] 
\nn \\
\nn
\eea
gives the number of muons in the i-th bin in the final muon energy
$E_\mu$, for a given neutrino energy $E_\nu$ (see App.~\ref{app:form}
for details).  In Fig.~\ref{fig:mudistr} we present a direct
comparison of ``golden'' and ``silver'' muons grouped in five energy
bins with $\Delta E_\mu = 10$ GeV, for a parent muon energy $\bar E_\mu
= 50$ GeV with input parameters $\bar \theta_{13} = 5^\circ$, $\bar
\delta = 60^\circ$. We consider here a near detector ($L = 732$ Km)
with a mass of 2 Kton and perfect reconstruction efficiency for both
channels. Notice that in Fig.~\ref{fig:mudistr} we have not included
quasi-elastic and resonance contributions to the $\nu N$ cross-section
that could enhance $\tau$ production at low neutrino energy.

The total number of events for these parameters are $N^g_{\mu^-} \sim
700$ and $N^s_{\mu^-} \sim 30$. The strong reduction in the number of
``silver'' muons with respect to the ``golden'' muons with the same
input parameters depends on the suppression due to the $BR (\tau \to
\mu)$ branching ratio and to the different $\nu N$ DIS cross-section
for muons and taus.

\begin{figure}[h!]
\begin{center}
\begin{tabular}{cc}
\hspace{-1cm} \epsfxsize7cm\epsffile{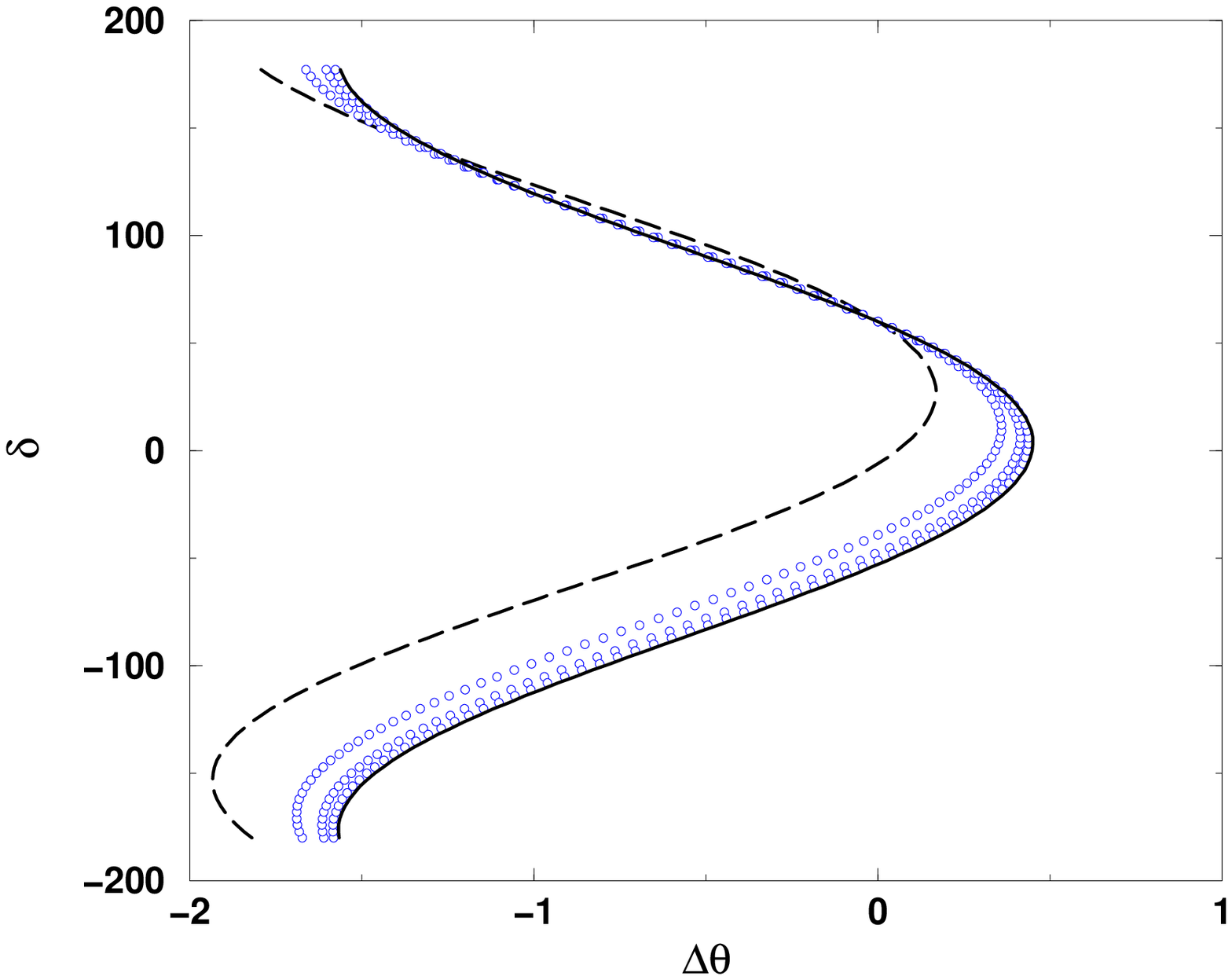} &
              \epsfxsize7cm\epsffile{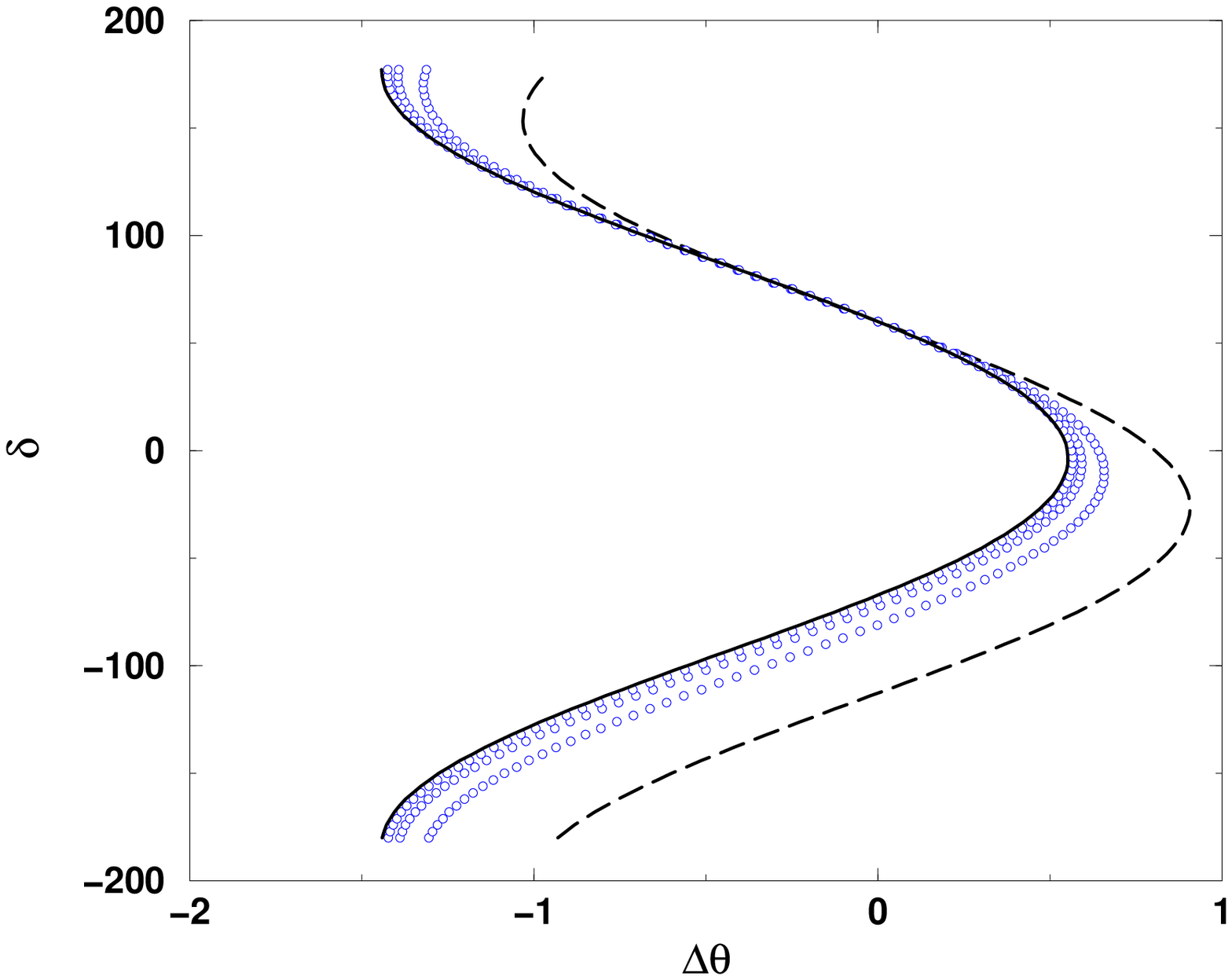}
\end{tabular}
\caption{\it Equal-number-of-event-curves in the ($\Delta \theta, \delta$) plane, 
for $E_\nu \in [5, 50] $ GeV and 
and $L = 732$ Km in the case of $\nu_e \to \nu_\tau$ oscillation
(neutrinos on the left, antineutrinos on the right), for $\bar \theta_{13} = 5^\circ$ 
and $\bar \delta = 60^\circ$.}
\label{fig:equiev2}
\end{center}
\end{figure}

Following the same procedure used to get the ``golden'' muons ENE
curves presented in Fig.~\ref{fig:equievents} we can compute ENE
curves for ``silver'' muons. These curves are reported in
Fig.~\ref{fig:equiev2} in the case of $L = 732$ Km.

\begin{figure}[h!]
\begin{center}
\epsfxsize10cm\epsffile{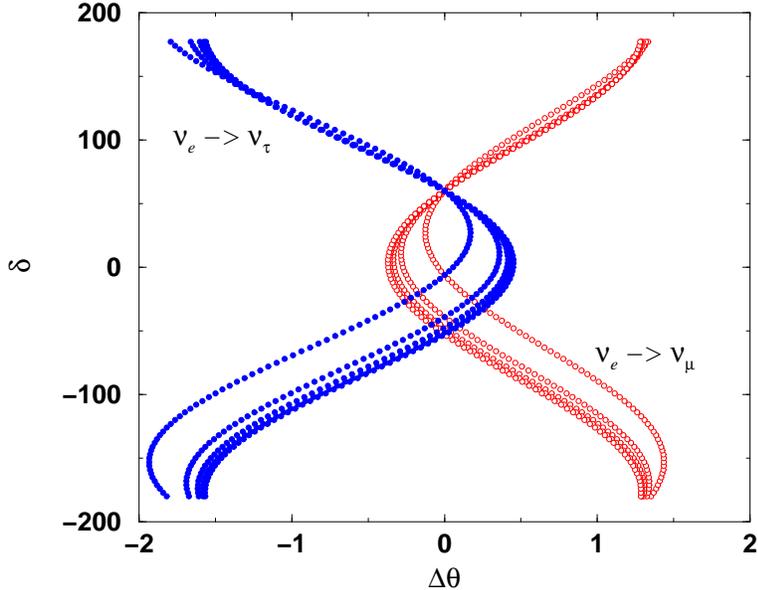}
\caption{\it Superposition of the equal-number-of-events curves for the transition 
$\nu_e \to \nu_\mu$ (light lines) and $\nu_e \to \nu_\tau$ (dark lines), 
for $L = 732$ Km and $\bar \theta_{13} = 5^\circ, \bar \delta = 60^\circ$.}
\label{fig:comb}
\end{center}
\end{figure}

In Fig.~\ref{fig:comb} we superimpose ``golden'' and ``silver'' ENE
curves for $L = 732$ Km and $\bar \theta_{13} = 5^\circ, \bar \delta =
60^\circ$.  Notice how, as it was expected from the equiprobability
curves analysis, the two sets of curves have opposite concavity in the
($\Delta \theta, \delta$) plane.  As for the equiprobability curves,
all lines met in the ``physical'' point.  Therefore, a combined
$\chi^2$ analysis of the two sets of data should present a well
defined global minimum around the ``physical'' region, whereas the
local minima situated in the three ``clone'' regions are considerably
raised with respect to what presented in the previous section where
only the ``golden'' muon signal was considered.

\section{Combination of ``golden'' and ``silver'' muon events} 
\label{sec:etauRES}

We follow the analysis in energy bins outlined in the previous section
and in \cite{Cervera:2000kp,Burguet-Castell:2001ez}: we produce a
theoretical data set ($N^g_{\mu^\pm}, N^s_{\mu^\pm}$) for fixed input
parameters $\bar \theta_{13}, \bar \delta$ and then we smear it as in
eq.~(\ref{eq:smearing}) to obtain an ``experimental'' data set
($n^g_{\mu^\pm}, n^s_{\mu^\pm}$). Finally, ``experimental'' data are
fitted and 68.5, 90 and 99 \% C.L. contours in the ($\theta_{13},
\delta$) plane are drawn.

\begin{figure}[h!]
\begin{center}
\begin{tabular}{cc}
\hspace{-1.5cm} \epsfxsize8.5cm\epsffile{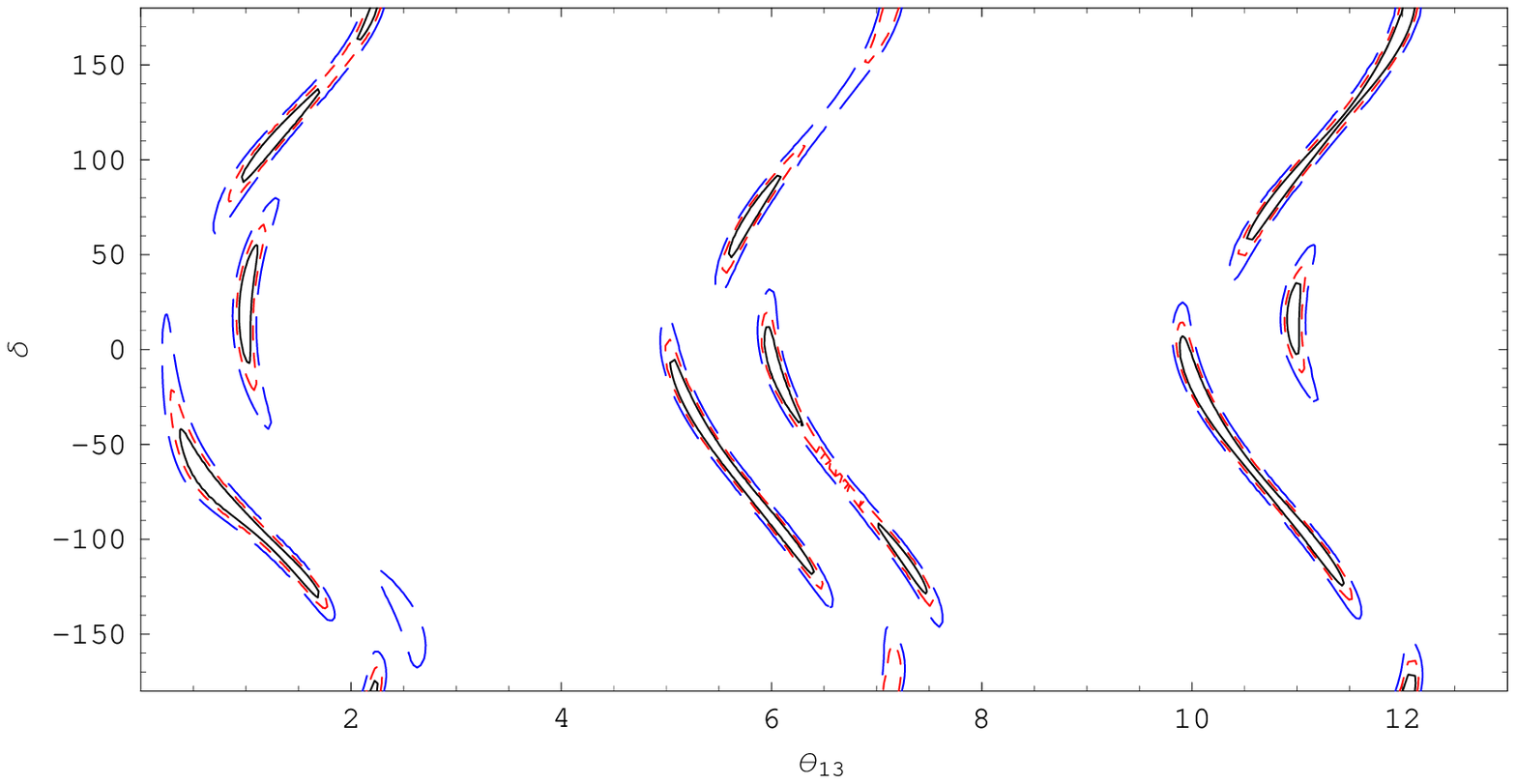} &
\epsfxsize8.5cm\epsffile{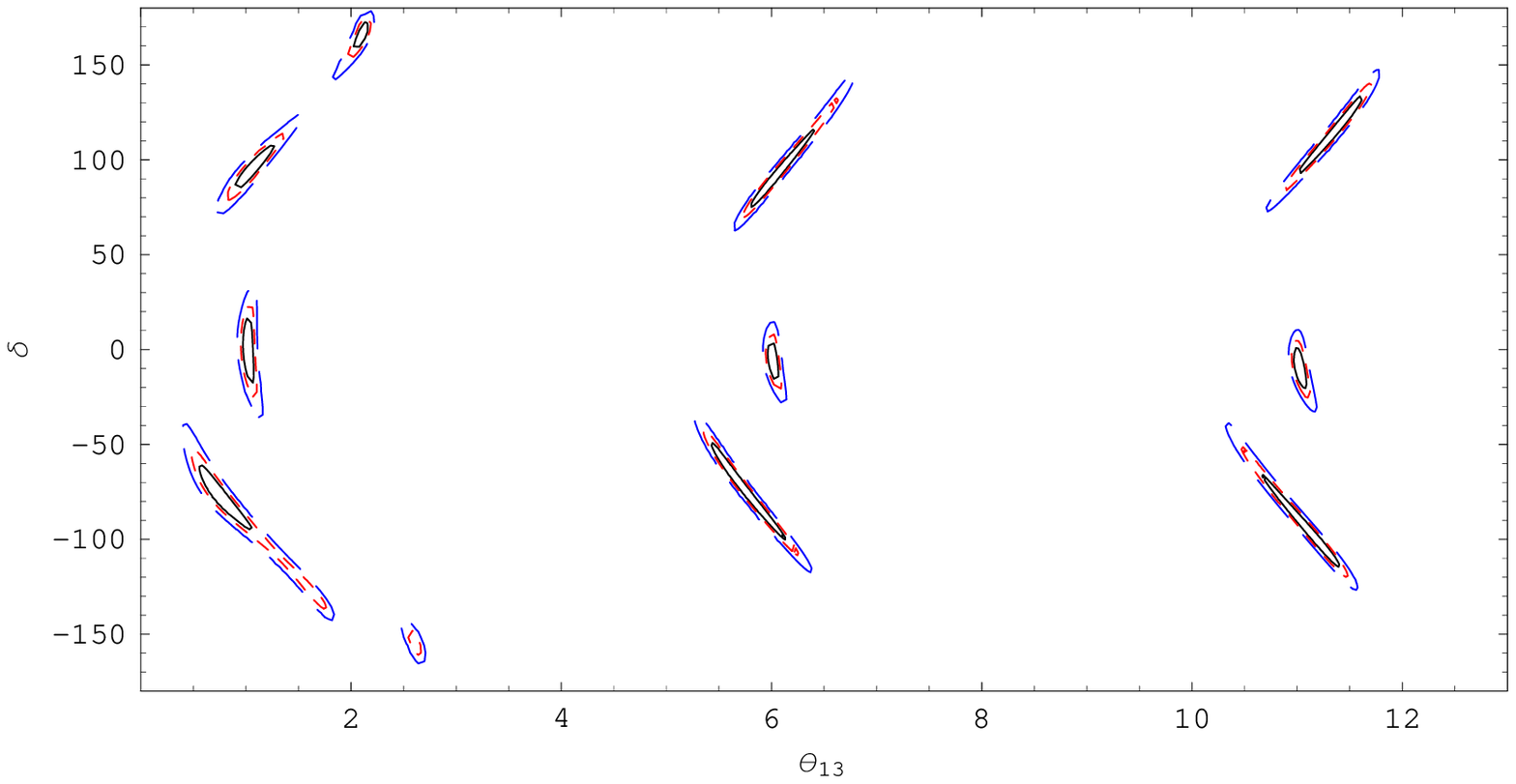} \\
\hspace{-1.5cm} \epsfxsize8.5cm\epsffile{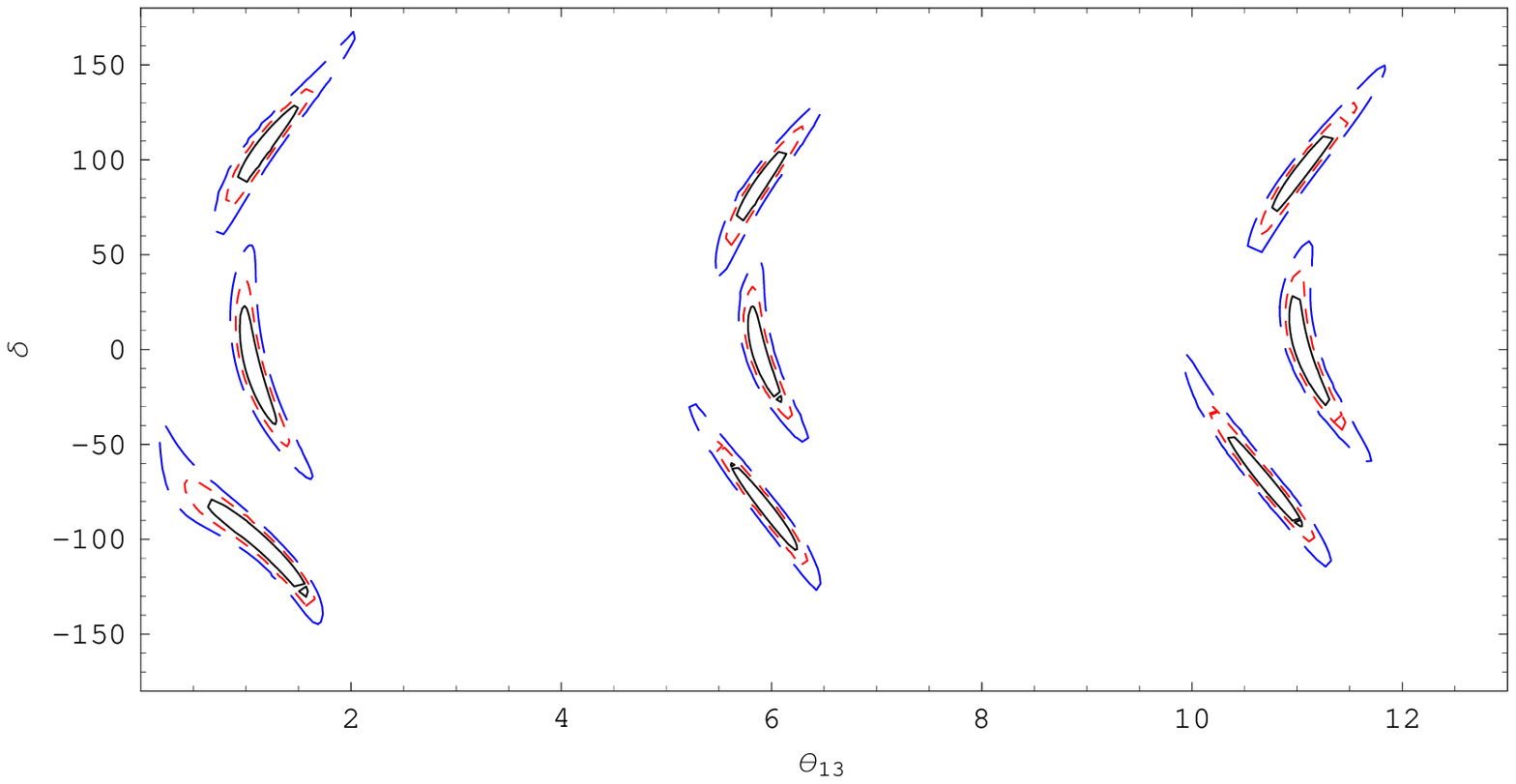} &
\epsfxsize8.5cm\epsffile{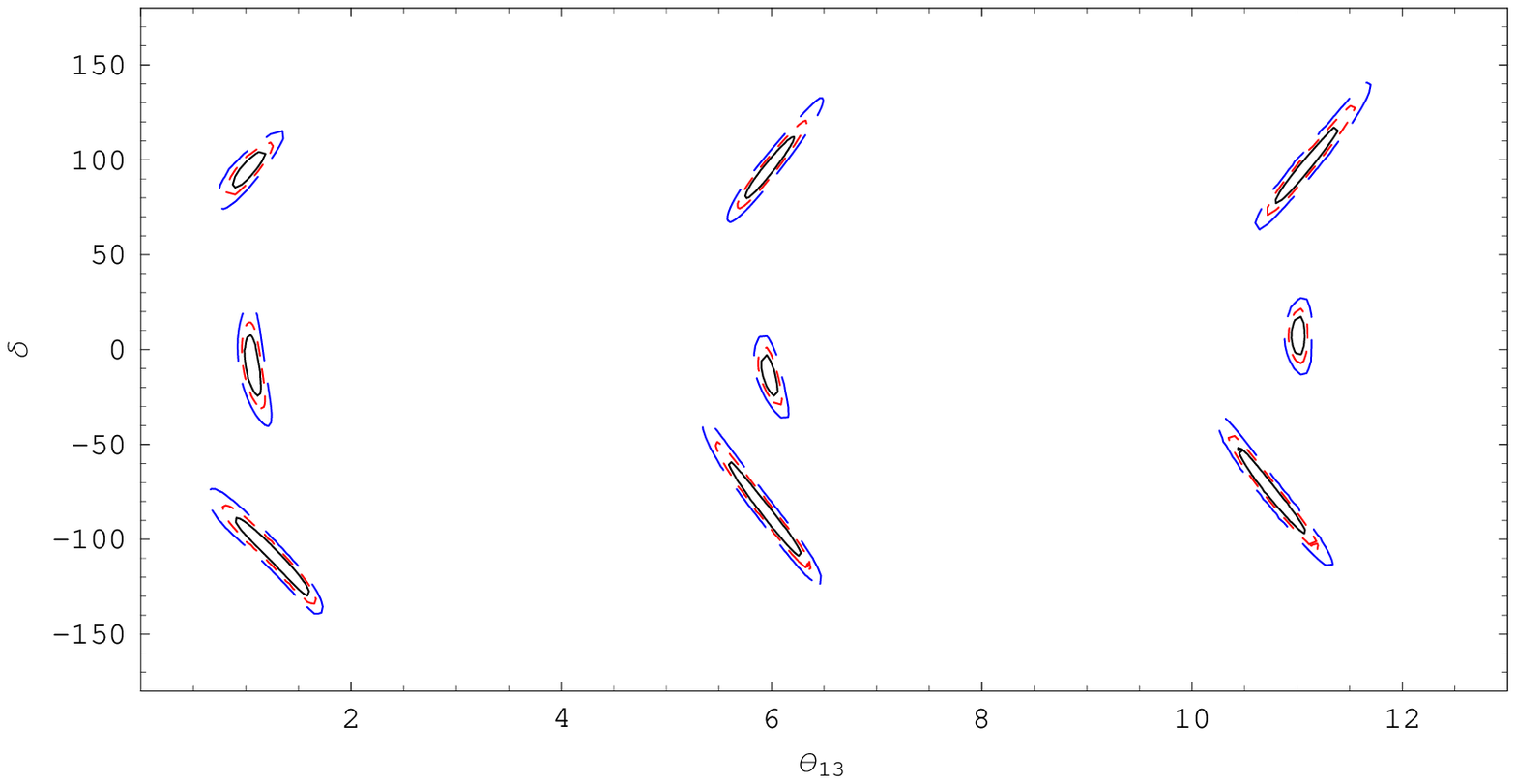} \\
\end{tabular}
\caption{ \it 68.5, 90 and 99 \% C.L. contours resulting from a $\chi^2$ fit of $\theta_{13}$
and $\delta$, for $\bar \theta_{13} = 1^\circ, 6^\circ$ and $11^\circ$, 
and $\bar \delta = -90^\circ, 0^\circ$ and $90^\circ$, for the combination of two iron
detectors (upper row) or one iron detector and one emulsion detector (lower row). 
Upper row: iron detector at $L = 732$ and 3000 Km:
a) $N^g_{\mu^+}$; b) $N^g_{\mu^+}$ and $N^g_{\mu^-}$;
Lower row: iron detector at $L = 3000$ Km and emulsion detector at $L = 732$ Km:
a) $N^g_{\mu^+}+ N^s_{\mu^+}$; b) $N^g_{\mu^+} + N^s_{\mu^+}$ and $N^g_{\mu^-}$.}
\label{fig:opiron}
\end{center}
\end{figure}

In Fig.~\ref{fig:opiron} we present the results of this analysis
comparing two different possibilities: in the upper row we combine two
realistic magnetized iron detectors at $L = 732$ and $L = 3000$ Km; in
the lower row we combine an ideal OPERA-like detector at $L = 732$ Km
and a realistic magnetized iron detector at $L = 3000$ Km.

First, we present our results for the combination of the two iron
detectors (where only ``golden'' muons, $N^g_{\mu^\pm}$, can be
used). We draw in each figure the contours for different input parameters: 
three values for $\bar \theta_{13} = 1^\circ, 6^\circ$ and $11^\circ$ and 
three values for the phase $\bar \delta = - 90^\circ, 0^\circ$ and $90^\circ$.
In each figure, therefore, fits to nine input parameter pairs 
($\bar \theta_{13},\bar \delta$) are shown: this has to be compared with 
Figs.~\ref{fig:th=1} and \ref{fig:th=5} where in each plot
the results of a fit to one single input parameter pair was presented. \\
On the left, only five years of data taking for $\mu^+$
circulating in the storage ring are considered.  
Notice that for any given input pair $\theta_{13}$ is always reconstructed within a
$2^\circ$ error; on the contrary, roughly any value for the CP-violating phase is
allowed. The situation is drastically improved on the right, where
five years of data taking for each muon polarization are
considered. In this case, the phase $\delta$ is reconstructed with a
precision of tens of degrees for all values of the input parameters.
Notice, however, how some ``clone'' region is still present at 90 \%
C.L. (e.g., for $\bar \theta_{13} = 1^\circ, \bar \delta = 90^\circ$ the
small region around $\theta_{13} = 2^\circ, \delta = 150^\circ$; 
for $\bar \theta_{13} = 1^\circ, \bar \delta = 0^\circ$ the
small region around $\theta_{13} = 2.5^\circ, \delta = - 150^\circ$; 
see also Fig.~\ref{fig:th=1}).
These results can be easily understood in terms of the
theoretical analysis of the equiprobability and equal-number-of-events
curves for the $\nu_e \to \nu_\mu$ channel of Sects. \ref{sec:emuprob}
and \ref{sec:emuENE}.

\begin{figure}[h!]
\begin{center}
\begin{tabular}{cc}
\hspace{-1.5cm} 
\epsfxsize8.5cm\epsffile{figure/nu_nub_opiron_3.ps} &
              \epsfxsize8.5cm\epsffile{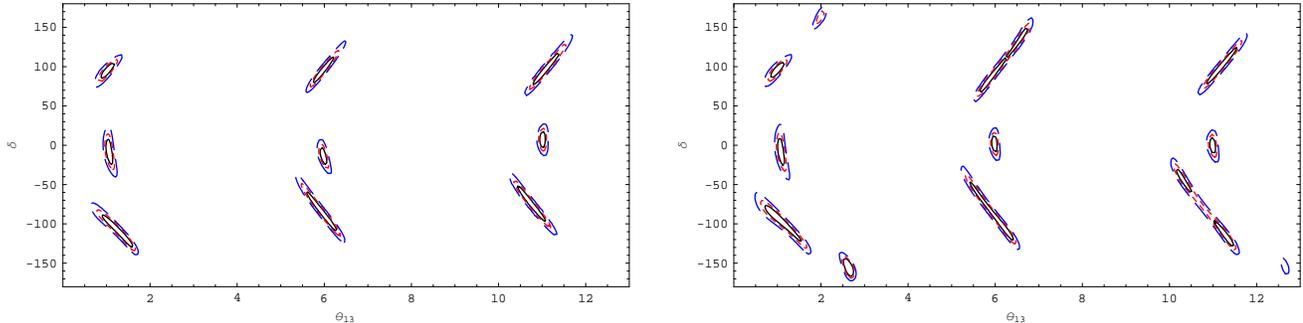}
\end{tabular}
\caption{ \it 68.5, 90 and 99 \% C.L. contours resulting from a $\chi^2$ fit of $\theta_{13}$
and $\delta$, for $\bar \theta_{13} = 1^\circ, 6^\circ$ and $11^\circ$, 
and $\bar \delta = -90^\circ, 0^\circ$ and $90^\circ$, for the combination of an iron 
detector at $L = 3000$ Km and an emulsion detector at $L = 732$ Km:
(left) both ``silver'' and ``golden'' muon events are taken into account (same
figure as in Fig.~\ref{fig:opiron}, lower right);
(right) only ``golden'' muon events are taken into account. 
Five years of data taking for both polarities 
in the distant detector and only five years in the $\mu^+$ polarity in the near detector
have been considered.}
\label{fig:notau}
\end{center}
\end{figure}

We present now results for the combination of a near emulsion detector (with
both ``golden'' and ``silver'' muons, $N^g_{\mu^\pm}$ and $N^s_{\mu^\pm}$)
and a not-so-far iron detector (with ``golden'' muons, only). On the left, again only 
five years of data taking for $\mu^+$ circulating in the storage ring are considered. 
Notice that a significant reduction in the reconstruction errors on the phase $\delta$ 
is already achieved. 
On the right, we simply add to the first five years of data taking for 
the $\mu^+$ polarity further five years for the opposite polarity in the iron 
detector, only. We have not included a further five year operational time 
for the emulsion detector to take into account the mass decrease due to the 
brick removal in the first five year period\footnote{This is a quite conservative assumption, 
being the expected number of bricks to be removed looking for ``silver'' muons
smaller than in the case of $\nu_\mu \to \nu_\tau$ oscillations
considered in the OPERA proposal, \cite{Guler:2000bd}.}. 
Notice, however, that a quite relevant improvement with respect to the one-polarity
two-detector types case (lower left) is achieved in the $\bar \delta$ reconstruction
error. More important, an improvement with respect to the two-polarities
one-detector type (upper right) can also be observed. In particular, the ``clone''
regions have completely disappeared, due to the combination of ``silver'' and ``golden''
muons with a different $\delta$-dependent oscillation probability.  
The effect of the inclusion of ``silver'' muons can be seen in Fig.~\ref{fig:notau}, 
where we present the two-polarities two-detector types combination 
(Fig.~\ref{fig:opiron}, lower right) compared with the same combination but
with only a ``golden'' muon signal. In this figure we can clearly see that ``clone''
regions are still present and that the near emulsion detector is too small to compete
with a 40 Kton iron detector located at the same distance down the neutrino source. 

We also studied the effect of a change in the longest baseline, moving the iron detector
from $L = 3000$ Km to $L = 2000$ Km, with no significant improvement. This was to 
be expected, since the former distance, $L = 3000$ Km has been found to be the optimal one
to measure the CP-violating phase for the considered set-up, 
\cite{Cervera:2000kp}-\cite{Rubbia:2001pk}.

We remind here that in \cite{Burguet-Castell:2001ez} it was shown that the optimal combination 
of two magnetized iron detectors is achieved for $L = 3000$ Km and $L = 7332$ Km: in that case, 
no ``clone'' region is present (see Fig.~6 of \cite{Burguet-Castell:2001ez}).
We prefer however to compare our results for the OPERA-like near detector and magnetized iron 
not-so-far detector option with the two magnetized iron detectors combination at the same
distances. 

Up to this moment, we considered an ideal lead-emulsion detector with
identification efficiency equal to 1 and no backgrounds. We should
consider instead a real detector with a refined analysis of
efficiencies and background for the two ``wrong-sign muons'' signals,
$\nu_e \to \nu_\mu$ and $\nu_e \to \nu_\tau$.  A detailed simulation
of the expected performance at an OPERA-like detector is under way
\cite{Autieroetal}. However, a preliminary study has been done
following the outline of the OPERA proposal \cite{Guler:2000bd} and of
the recent progress report \cite{Guler:2001hz}.

We first consider ``golden'' muons. In this case, it seems reasonable
to consider a 90 \% average efficiency in the neutrino energy range
$E_\nu \in [5,50]$ GeV.  The most relevant background to this channel
are ``right-sign'' muons with a wrong charge assignment: with no
improvement with respect to the present OPERA proposal, a level of
$10^{-3}$ of charge misidentification is achieved (a level of
$10^{-6}$ was envisaged for the magnetized iron detector, see
\cite{Cervera:2000vy}).  On the other side, the dominant background at
the magnetized iron detector coming from charmed mesons decay with no
``right-sign'' muon identification is here under control by looking
for the decay vertex.

Regarding ``silver'' muons, we have an unavoidable important reduction in the average 
efficiency, mainly due to the detector design: the $\tau$ decay vertex cannot 
be observed if it occurs inside the lead plates. 
An average efficiency for ``silver'' muons of 25 \% only is achieved 
in the neutrino energy range $E_\nu \in [5,50]$ GeV, \cite{Guler:2000bd}.
\begin{figure}[h!]
\begin{center}
\epsfxsize10cm\epsffile{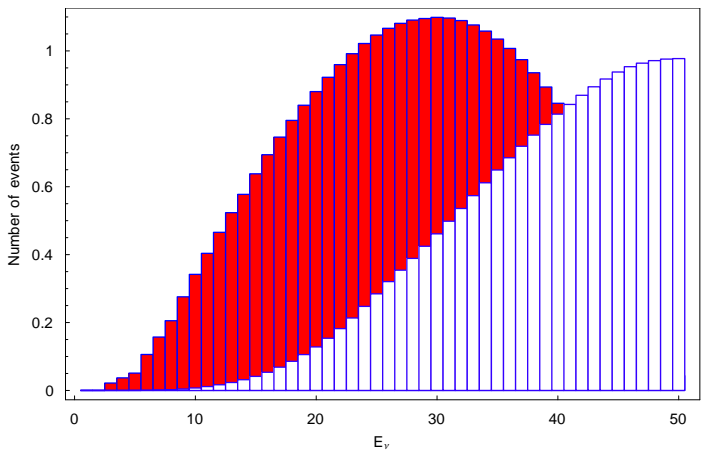}
\caption{\it Charmed mesons decay background and ``silver'' muons signal
(shaded) for $\bar \theta_{13} = 11^\circ$ and $\bar \delta = 90^\circ$
in the OPERA-like emulsion detector as a function of the reconstructed 
neutrino energy (for 1 GeV energy bins). 
}
\label{fig:back}
\end{center}
\end{figure}
In the case of ``silver'' muon events, the dominant background is the
standard charmed mesons decay with no ``right-sign'' muon identification.
Sub-dominant backgrounds coming from $\pi$ and $K$ mesons decay can be kept at
a reasonable level by imposing kinematical cuts as described in \cite{Guler:2000bd}. 
As an illustrative example, in Fig.~\ref{fig:back} we present
the charmed mesons decay background and the ``silver'' muon signal for 
$\bar \theta_{13} = 11^\circ$ and $\bar \delta = 90^\circ$ as a function of the 
neutrino energy (for 1 GeV energy bins). Notice how the ``silver'' muons signal
peaks at lower energy with respect to the background, thus allowing us to use
the $\nu_e \to \nu_\tau$ oscillation to improve the ($\theta_{13}, \delta$)
reconstruction.

\begin{figure}[h!]
\begin{center}
\epsfxsize12cm\epsffile{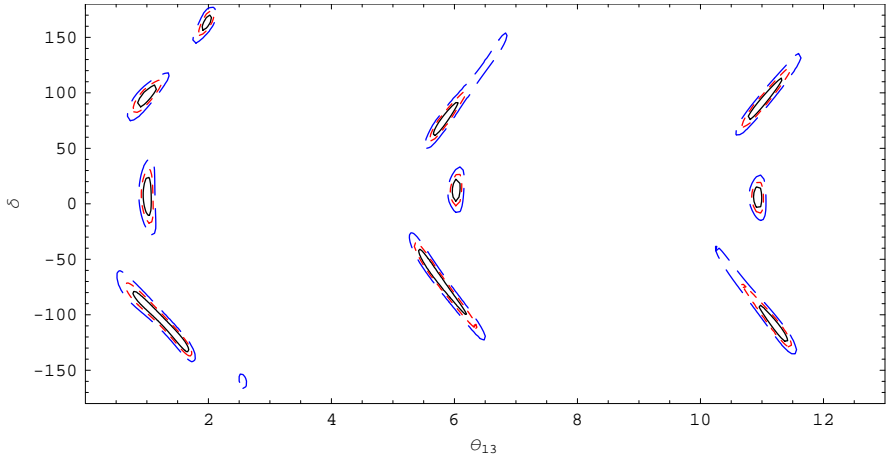}
\caption{ \it 68.5, 90 and 99 \% C.L. contours resulting from a $\chi^2$ fit of $\theta_{13}$
and $\delta$, for $\bar \theta_{13} = 1^\circ, 6^\circ$ and $11^\circ$, 
and $\bar \delta = -90^\circ, 0^\circ$ and $90^\circ$, for the combination of an iron 
detector at $L = 3000$ Km (with ``golden'' muon events) and an emulsion detector at $L = 732$ Km
(with ``golden'' and ``silver'' muon events). Realistic estimates for
efficiencies and backgrounds at both the iron and the emulsion detector have been taken
into account. 
Five years of data taking for both polarities in the distant detector and only five years 
in the $\mu^+$ polarity in the near detector have been considered.}
\label{fig:operareal}
\end{center}
\end{figure}

Eventually, we present in Fig.~\ref{fig:operareal} the results for the combination of an 
iron detector at $L = 3000$ Km and of an emulsion detector at $L = 732$ Km with
reasonable estimates for the reconstruction efficiencies for ``golden'' 
and ``silver'' muons and for the corresponding backgrounds. We can see that
the results obtained with the two-detector types combination strongly resembles
those obtained with the reference two iron detectors set-up (Fig.~\ref{fig:opiron}, upper right). 
In particular, we notice a general improvement at the 68 \% C.L. but we still observe 
``clone'' regions for small values of $\bar \theta_{13}$. 

Although the results of Fig.~\ref{fig:operareal} do not indicate
a strong improvement with respect to the reference two iron detectors
set-up, two comments are in order: first, we have been considering up
to now an OPERA-like detector with the characteristics of the detector
described in \cite{Guler:2000bd} which is currently under construction
and should be operational by year 2006. We therefore study the effects 
on the physics reach of foreseeable improvements in the following.
Secondly, it seems to us that the combination of two different detector 
types with different characteristics and systematics could be of the utmost
importance in the project of a facility for precision measurements of
the leptonic mixing matrix such as the Neutrino Factory.

In Tab.~\ref{tab:impro} we schematically present which are the main
advantages of an increase in the emulsion detector mass, or of a
reduction of the main backgrounds, considering the presence or absence
of ``clone'' regions and the maximal error in $\delta$ that can be
achieved.  The main background for ``silver'' muon events $B_{silver}$
(originating from charmed mesons decay) is normalized to the value
that can be deduced on the basis of the OPERA proposal.  In this case
we do not consider that significant improvements can be achieved and
we restrict ourselves to a 10 to 20 \% reduction.  We believe that, on
the other hand, the main background for ``golden'' muon events
$B_{golden}$ (originating from charge misidentification) could be
significantly reduced: we therefore present results for a $10^{-3}$
(the predicted level in the OPERA proposal) and for a $10^{-4}$ charge
misidentification. Finally, having in mind that in the last years
the scanning power increased by about a factor 10 every two years, we
consider as realistic the assumption that, by the time the Neutrino
factory will become operational, we will be able to scan a larger
number of events, say a factor two. Therefore, we studied the physical
reach of the Neutrino factory-detectors set-up as a function of the
lead-emulsion mass, considering a linear increase in the mass from 2 Kton to 4 Kton. 

\begin{table}[t]
\centering
\begin{tabular}{||c|c|c|c|c||}
\hline\hline
 Mass (Kton) & $B_{silver}$  & $B_{golden}$ & Clone regions & $\delta_{min},\delta_{max}$
at 99 \% C.L. \\
\hline \hline
  & 1.0 & $10^{-3}$ & At 68 \% C.L. & - \\
  & 0.9 & $10^{-3}$ & At 68 \% C.L. & - \\
  & 0.8 & $10^{-3}$ & At 68 \% C.L. & - \\
2 & 1.0 & $10^{-4}$ & At 68 \% C.L. & - \\
  & 0.9 & $10^{-4}$ & At 90 \% C.L. & - \\
  & 0.8 & $10^{-4}$ & At 90 \% C.L. & - \\
\hline \hline
  & 1.0 & $10^{-3}$ & At 99 \% C.L. & - \\
  & 0.9 & $10^{-3}$ & At 99 \% C.L. & - \\
  & 0.8 & $10^{-3}$ & At 99 \% C.L. & - \\
3 & 1.0 & $10^{-4}$ & No & $50^\circ$ - $100^\circ$ \\
  & 0.9 & $10^{-4}$ & No & $50^\circ$ - $110^\circ$ \\
  & 0.8 & $10^{-4}$ & No & $50^\circ$ - $110^\circ$ \\
\hline \hline
  & 1.0 & $10^{-3}$ & No & $55^\circ$ - $105^\circ$ \\
  & 0.9 & $10^{-3}$ & No & $55^\circ$ - $105^\circ$ \\
  & 0.8 & $10^{-3}$ & No & $55^\circ$ - $105^\circ$ \\
4 & 1.0 & $10^{-4}$ & No & $60^\circ$ - $110^\circ$ \\
  & 0.9 & $10^{-4}$ & No & $60^\circ$ - $110^\circ$ \\
  & 0.8 & $10^{-4}$ & No & $70^\circ$ - $110^\circ$ \\
\hline \hline
\end{tabular}
\caption{\it Effects of a mass increase or of a ``silver'' ($B_{silver}$) 
or ``golden'' ($B_{golden}$) main background reduction for the emulsion detector, 
for $\bar \theta_{13} = 1^\circ$ and $\bar \delta = 90^\circ$. Whenever a ``clone''
region is present, no overall error on $\delta$ is given.}
\label{tab:impro}
\end{table} 

The outcome of this analysis is the following: 
\begin{itemize}
\item
``Clone'' regions disappear increasing the emulsion mass. For a 4 Kton detector
no ambiguity is present. An improvement in the charge identification (i.e., a reduction
of ``golden'' muons main background) implies generally a decrease in the statistical 
significance of ``clone'' regions.
\item
When no ``clone'' regions are observed, a reduction of ``golden'' muons main background
induces a shift in the 99 \% C.L. allowed region for $\delta$ towards higher values. 
In the optimal case we observe a gaussian-distributed region with $\Delta \delta = \pm 20^\circ$.
\item
A 10 to 20 \% reduction of the ``silver'' muons main background (charmed mesons decay) 
does not seem to improve significantly the previous results. 
\end{itemize}

We remind that this table should be interpreted as an indication of which kind of 
improvement in the emulsion detector helps more to improve the physics reach of the
envisaged two detector types combination. 

\section{Conclusions}
\label{sec:concl}

It was previously shown \cite{Cervera:2000kp} that the ``golden'' wrong-sign muon
signal at the Neutrino Factory can be extremely useful to measure simultaneously
two of the parameters of the PMNS leptonic mixing matrix, $\theta_{13}$ and $\delta$. 
In \cite{Burguet-Castell:2001ez} was first noticed that degenerate regions in the 
($\theta_{13}, \delta$) parameter space occur in many cases, severely reducing 
the Neutrino Factory sensitivity to the CP-violating phase $\delta$. 

In the first part of the paper we describe how the ($\theta_{13}, \delta$) ambiguity
arises in $\nu_e \to \nu_\mu$ oscillation due to the equiprobability curves 
in the ($\theta_{13},\delta$) plane at fixed neutrino energy. 
We then extend our analysis showing how the same phenomenon
can be observed in a real experiment: equal-number-of-events (ENE) curves 
for any given neutrino energy bin appears and their intersections in the 
($\theta_{13},\delta$) plane explain how and where do ``clone'' regions
arise. Our analysis is subsequently compared with the results of simulations 
with a realistic magnetized iron detector, already studied in previous papers.

In the second part of the paper, we propose the use of $\nu_e \to
\nu_\tau$ oscillation to solve the ($\theta_{13}, \delta$) ambiguity
problem. We apply the arguments presented in the case of $\nu_e \to
\nu_\mu$ and we show how $\nu_e \to \nu_\tau$ actually give a
complementary information that can be used to improve the measurement
of the phase $\delta$.  However, to take full advantage of this
``silver'' channel we should use a detector of a different kind to
distinguish muons originated from $\tau$ decay from the so-called
``golden'' muons.  In the last part of the paper we therefore perform
simulations using the combination of a realistic magnetized iron
detector at $L = 3000$ Km and of an OPERA-like lead-emulsion
detector located at $L = 732$ Km from the Neutrino Factory. 
The $\tau$ decay vertex recognition in the emulsion detector
allow us to separate ``silver'' from ``golden'' muon events and
to strongly reduce the effect of the charmed mesons decay background.

We first use an ideal lead-emulsion detector with the mass
currently envisaged by the OPERA Collaboration (2 Kton), perfect
reconstruction efficiency and no background, showing how the ``clone''
regions disappear for any value of $\theta_{13} \geq 1^\circ$.
Eventually, we consider a realistic estimate of the reconstruction
efficiency and of the main backgrounds both for ``golden'' and
``silver'' muon events at the emulsion detector, following the outline
of the OPERA proposal \cite{Guler:2000bd}. In this case, ``clone''
regions for $\theta_{13} \simeq 1^\circ$ do appear and our results are
similar to those that we obtain for a combination of two realistic
magnetized iron detectors with $L = 732$ and $L = 3000$
baselines. However, we stress here that the OPERA-like 2 Kton
lead-emulsion detector is under construction and should be operational
by mid 2006 to exploit the CNGS beam. We can therefore study
how possible improvements of the detector (from 2006 till the time the 
Neutrino factory will become operational) affect the results of
our analysis. In particular we remind that a moderate scaling
in the lead-emulsion detector mass could indeed be seriously
taken into account (due to the rapid increase in the emulsion 
scanning power). The outcome of this study is that an increase in the
emulsion mass from 2 to 4 Kton eliminates completely the ``clone''
regions for $\theta_{13} \geq 1^\circ$. On the other
hand, a moderate increase in the main background rejection
does not seem to improve in a significant way the previous results. 

As a final comment we believe that the two-detector types combination 
should be investigated further, even in the case of results comparable
to those of a two-baselines magnetized iron detector combination,
since having detectors with different
characteristics and systematics can be extremely helpful to ameliorate
the $\delta$ measurement.

\section*{Acknowledgements}
We are deeply indebted with Paolo Lipari and Maurizio Lusignoli for exhaustive
discussions on many different aspects of this paper. We acknowledge useful discussions
with L. Di Lella, M.B. Gavela, J.J. Gomez-Cadenas, P. Hernandez, L. Ludovici, 
Carlos Pe\~na-Garay, S. Rigolin, G. Rosa, P. Strolin, D. Zardetto and P. Zucchelli.

\newpage

\appendix
\section{Perturbative expansion in $\Delta \theta$}
\label{app:dtheta}

If $\theta_{13}$ is large enough, we can expand eqs.~(\ref{eq:functions})
and (\ref{eq:equi2}) in powers of $\Delta \theta$ for any value of $\Delta \theta$ 
in the allowed region of Fig.~\ref{fig:dthetalim}.

\noindent At first order in $\Delta \theta$, 
\be
\label{eq:fgapprox}
\left \{ \begin{array}{cccc}
        f (\theta_{13}, \bar \theta_{13} ) &\simeq& - \Delta \theta
        \, \tilde f(\bar \theta_{13}) + {\cal O} (\Delta \theta^2)
& \simeq
- 4 \Delta \theta \, \frac{\cos (2 \bar \theta_{13})}{\sin (\bar \theta_{13})} 
+ {\cal O} (\Delta \theta^2) \, , \\
\\
        g (\theta_{13}, \bar \theta_{13} ) &\simeq& 1 - \Delta \theta 
        \, \tilde g(\bar \theta_{13}) + {\cal O} (\Delta \theta^2)
& \simeq
1 - \Delta \theta \, \frac{3 \cos( 2 \bar \theta_{13})- 1 }{\sin ( 2 \bar \theta_{13}) }
+ {\cal O} (\Delta \theta^2) \, ,
         \end{array} \right .
\ee
where, for $\bar \theta_{13} \in [0^\circ, 13^\circ]$, $\tilde f$ and $\tilde g$ are
positive functions of $\bar \theta_{13}$. \\
For neutrinos we get
\be
\delta_+ = \frac{\Delta_{atm} L}{2} \pm \left \{ 
  \left ( \bar \delta - \frac{\Delta_{atm} L}{2} \right ) -
  \Delta \theta \, \frac{ 4 R_+ \tilde f (\bar \theta_{13}) +
  \cos \left (\bar \delta - \frac{\Delta_{atm} L}{2} \right ) \tilde g (\bar \theta_{13})
}{
|\sin \left (\bar \delta - \frac{\Delta_{atm} L}{2} \right )|}
\right \}
\label{eq:deltapapprx}
\ee
and for antineutrinos we get
\be
\delta_- = - \frac{\Delta_{atm} L}{2} \pm \left \{ 
  \left ( \bar \delta + \frac{\Delta_{atm} L}{2} \right ) -
  \Delta \theta \, \frac{ 4 R_- \tilde f (\bar \theta_{13}) +
  \cos \left (\bar \delta + \frac{\Delta_{atm} L}{2} \right ) \tilde g (\bar \theta_{13})
}{
|\sin \left (\bar \delta + \frac{\Delta_{atm} L}{2} \right )|}
\right \} \, .
\label{eq:deltamapprx}
\ee
Notice that from eq.~(\ref{eq:functions}), for $\Delta \theta = 0$, 
we get $\delta_\pm = \bar \delta$ or 
$\delta_\pm = - \bar \delta \pm \Delta_{atm} L$. For $\Delta \theta \neq 0$, 
we get the two branches of the equiprobability curves. 

If we turn on the $\Delta \theta$ correction at first order, 
the results of Fig.~\ref{fig:equiprob} are easily understood with the 
following argument: 
consider two different equiprobability curves for neutrinos, 
for two different values of the neutrino 
energy $E_1$ and $E_2$, $E_2 \ge E_1$. 
The intersections between the two curves are defined by 
\bea
\delta_+(E_1) &=& \delta_+ (E_2) \, . \nn \\
\nn
\eea
Solving for $\Delta \theta$ we get two solutions:
\be
\left \{ 
\begin{array}{lll}
\Delta \theta & = & 0 \, , \\
&& \\
\Delta \theta & = & L \, \frac{\Delta_{atm}(E_1) - \Delta_{atm}(E_2)}{H(E_2) - H(E_1)}
\, , 
\end{array}
\right .
\ee
where
\bea
H (E) = \frac{ 4 R_+ (E) \tilde f (\bar \theta_{13}) +
  \cos \left (\bar \delta + \frac{\Delta_{atm}(E) L}{2} \right ) 
\tilde g (\bar \theta_{13})
}{
|\sin \left (\bar \delta + \frac{\Delta_{atm} (E) L}{2} \right )|} \, . \nn \\
\nn
\eea
For $\Delta \theta = 0$, we get $\delta = \bar \delta$.
On the other hand, the second intersection gets an ${\cal O} (\Delta \theta)$
correction:
\be
\delta = - \bar \delta + L \Delta_{atm} (E_1) + \Delta \theta H (E_1) \, .
\ee
Being $H (E)$ and (for the considered parameters and baselines)
the second solution for $\Delta \theta$ both positive quantities, 
the second intersection between any two equiprobability curves for neutrinos
will be displaced towards negative values of $\delta$ and positive values
of $\Delta \theta$. This is precisely what observed in the upper row of 
Fig.~\ref{fig:equiprob}. 
For the considered set of input parameters, baselines and neutrino energy range
the same argument applies to the antineutrino equiprobability curves also,
as it can be seen in the lower row of Fig.~\ref{fig:equiprob}.

\section{Formulae for $\tau$ decay and $\nu_\tau N$ cross-section}
\label{app:form}

Grouping events in bins of the final muon energy $E_\mu$, with the size of the 
energy bin depending on the energy resolution $\Delta E_\mu$ of the considered detector, 
the number of ``silver'' muon events in the i-th energy bin for the input pair 
($\bar \theta_{13}, \bar \delta$) and for a parent muon energy $\bar E_\mu$ is:
\bea
N^s_{\mu^\mp} (\bar \theta_{13}, \bar \delta)
&=& BR(\tau \to \mu) \, 
   \left \{ \left [ 
      \frac{d N_{\mu^\mp} (E_\mu, E_\tau) }{d E_\mu}                    
                           \, \otimes \,
      \frac{d \sigma_{\nu_\tau (\bar \nu_\tau)} (E_\tau, E_\nu) }{d E_\tau}
           \right ] \right .\nn \\
& & \left .  \qquad             \, \otimes \,
                 P^\pm_{e\tau} (E_\nu, \bar \theta_{13}, \bar \delta)           
                           \, \otimes \,
                 \frac{d \Phi_{\nu_e (\bar \nu_e) } (E_\nu, \bar E_\mu)}{d E_\nu} 
                      \right \}_{E_i}^{E_i + \Delta E_\mu} \, , \nn
\eea
where $\otimes$ stands for a convolution integral on the intermediate energy.

Several comments are in order: 
\begin{itemize}
\item 
The differential neutrino flux $d \Phi_{\nu_e (\bar \nu_e)}(E_\nu, \bar E_\mu)/d E_\nu$ 
has been given in eq.~(\ref{fluxes}). 
\item
The $\nu_e \to \nu_\tau$ oscillation probability
has been given in eq.~(\ref{eq:etau}).
\item {\bf The $\nu_\tau N$ cross-section} \\
The (anti)neutrino-nucleon differential cross sections on an isoscalar target, 
defined as the average of the differential cross-sections on proton and neutron, 
can be divided in three components: the elastic, the quasi-elastic and the DIS cross-section. 
We refer to \cite{Zardetto} for a detailed discussion on the different contributions to these 
components. The DIS cross-section can be expressed in terms of the isoscalar 
structure functions $F_i (x)$ ($i = 1, \dots, 5$): 
\bea 
\frac{d^2 \sigma^{\nu (\bar \nu)}}{dx dy} &=& \frac{G_F^2 M E_\nu}{\pi} \, 
\left \{
y \left ( x y + \frac{m_l^2}{2 M E_\nu} \right ) F_1 (x) +
\left ( 1 - y - \frac{M x y}{2 E_\nu} - \frac{m_l^2}{4 E_\nu^2} \right ) F_2 (x) 
\right . \nn \\
& \pm &
\left [ x y \left ( 1 - \frac{y}{2} \right ) - y \frac{m_l^2}{4 M E_\nu} \right ] F_3 (x)
+ \left ( x y \frac{m_l^2}{2 M E_\nu} + \frac{m_l^4}{4 M^2 E_\nu^2} \right ) F_4 (x) \nn \\
&-& \left . \frac{m_l^2}{2 M E_\nu} F_5 (x) \right \} \, , 
\label{eq:DIS}
\eea
where $M$ is the nucleon mass and $m_l$ the charged lepton mass; $x = Q^2 / 2 M E_\nu y$
is the DIS Bjorken variable and $y = 1 - E_l / E_\nu$ depends on the fraction of neutrino
energy carried away by the charged lepton. Notice that the $x,y$ variables defined in
this Appendix should be interpreted as {\it local variables}: outside the Appendix
the same letters may represent different objects (we are simply following the standard 
notation for the DIS cross-section). 

Making use of the Callan-Gross relation, $2 x F_1 (x) = F_2 (x)$, and of the 
Albright-Jarlskog relations \cite{Albright:1975ts}, $F_4 (x) = 0, x F_5 (x) = F_2 (x)$, 
eq.~(\ref{eq:DIS}) depends on two independent structure functions, only. 
Using the quark parton model (see \cite{Paschos:2001np} and refs. therein, for example) 
the isoscalar structure functions $F_2 (x)$ and $x F_3 (x)$ are:
\be
\left \{ 
\begin{array}{lll}
F_2^\nu (x) & = & x \left [ (u + d + 2 s) + ( \bar u + \bar d + 2 \bar c) \right ]
\, , \\
F_2^{\bar \nu} (x) & = & x \left [ (u + d + 2 c) + ( \bar u + \bar d + 2 \bar s) \right ]
\, , \\
&& \\
x F_3^\nu (x) & = & x \left [ (u + d + 2 s) - ( \bar u + \bar d + 2 \bar c) \right ]
\, , \\
x F_3^{\bar \nu} (x) & = & x \left [ (u + d + 2 c) - ( \bar u + \bar d + 2 \bar s) \right ]
\, .
\end{array}
\right .
\ee
Eventually, the structure functions, given in terms of the parton distribution 
at a definite value of $Q^2$ \cite{Martin:2001es}, 
must be evolved according to the DGLAP equations
to compute the $\nu_\tau N$ cross-section for a given neutrino energy.

\begin{figure}[h!]
\begin{center}
\hspace{-1cm} \epsfxsize8cm\epsffile{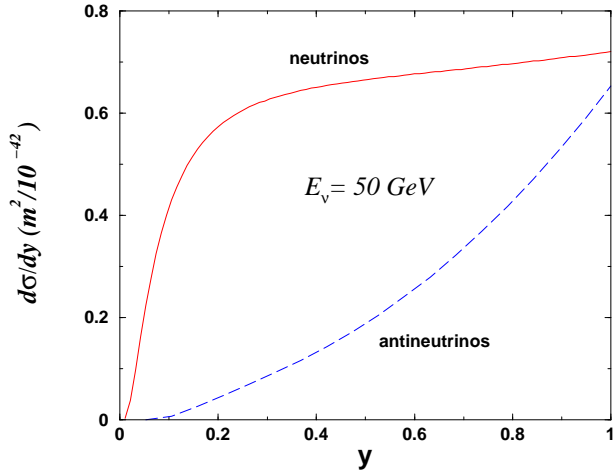} 
\caption{\it Neutrino and antineutrino -- nucleon differential cross
section as a function of $y$ for $E_\nu = \bar E_\mu$.}
\label{fig:cross}
\end{center}
\end{figure}

In Fig.~\ref{fig:cross} we show the dependence of the $d\sigma/dy$ on $y$ 
for a fixed value of $E_\nu$ (in this case, $E_\mu = \bar E_\mu$).
Notice the strong suppression of the antineutrino cross-section with respect to 
the neutrino one for high and intermediate $\tau$ energy ($y \to 0$).
The integration limits for the $(x,y)$ variables are reported in Fig.~\ref{fig:xbounds} 
for different neutrino energies \cite{Zucchelli}.

\begin{figure}[h!]
\begin{center}
\hspace{-1cm} \epsfxsize13cm\epsffile{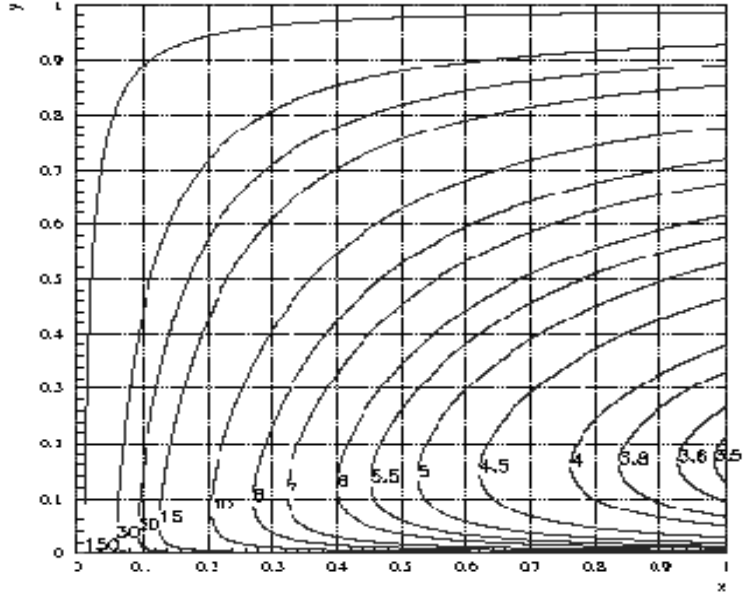} 
\vspace{0.5cm}
\caption{\it Kinematical bounds in the $x,y$ plane for different neutrino energies, 
\cite{Zucchelli}.}
\label{fig:xbounds}
\end{center}
\end{figure}

\item {\bf The $\tau$ decay rate} \\
The muon distribution is given, in the $\tau$ rest frame, by the following expression: 
\be
\frac{d^2 N_{\mu^\pm}}{d x^\prime d \cos \theta} = \frac{1}{2} 
\left [ f_0 (x^\prime,R) \mp {\cal P_\tau} f_1 (x^\prime,R) \cos \theta \right ]
\ee
where $R = m_\mu / m_\tau$, ${\cal P_\tau}$ is the average $\tau$ polarization along 
the $\tau$ direction in the laboratory frame, $x^\prime = 2 E_\nu/m_\tau$ 
(not to be confused with the Bjorken $x$ variable defined for the cross-section) 
and $\theta$ is the angle between the muon momentum vector and the $\tau$ spin direction.
The two $f$-functions are: 
\bea
f_0(x^\prime,R) &=& 2 \sqrt{x^{\prime 2} - 4 R^2} \, 
          \left [ - 4 R^2 + 3 (1+ R^2) x^\prime - 2 x^{\prime 2} \right ] \\
f_1(x^\prime,R) &=& 2 \left ( x^{\prime 2} - 4 R^2 \right )  \, 
          \left [ 1 + 3 R^2 - 2 x^\prime \right ]
\eea
and they reduce to the standard $f_0(x^\prime)$ and $f_1(x^\prime)$ functions for 
the $\nu_\mu$ and $e$ flux in the $R \to 0$ limit \cite{gaisser}.
The boost in the laboratory frame is given by the following relations: 
\bea
y' & = & \frac{x'}{2} + \frac{\beta}{2} \left[x'^2- 4R^2 \right]^{1/2} cos \,\theta 
\cr \cr
cos \,\theta^\prime  & = & \frac{\beta\,x'+ \left(x'^2- 4 R^2 \right)^{1/2} cos \,\theta}
{\left[(x'+\beta (x'^2- 4 R^2)^{1/2}\,cos \,\theta)-4(1-\beta^2) R^2 \right]^{1/2}} 
\label{relations}
\eea
where $y^\prime = E_\mu/E_\tau$ and $\theta^\prime$ is the angle boosted in the lab frame.
\end{itemize}

\end{document}